\documentclass{aastex}

\slugcomment{submitted to ApJ}

\shorttitle{CMB Polarization $E$ and $B$ Modes}
\shortauthors{Chiueh and Ma}

\begin{document}


\title{The Annulus-Filtered $E$ and $B$ Modes in CMBR Polarization}


\author{Tzihong Chiueh and Cheng-Jiun Ma}
\affil{Physics Department, National Taiwan University,
    Taipei, Taiwan}
\affil{Institute of Astronomy and Astrophysics, Academia Sinica, Taipei,
Taiwan}




\begin{abstract}
Although it has previously been recognized that the CMB
polarization measurement on a sky ring is able to
separate the $B$ and $E$-mode patterns,  
we show that the rotational symmetry
of the CMB polarization measurement
is indeed unique, in that it enables the  
$B$ and $E$ modes to be separately measured
in a simple and clean manner.  The separation of $B$ and $E$ modes 
can in principle be achieved even with a
single-polarization detector of arbitrary beam pattern.
Based on this premise, a specific observing
strategy is suggested, where
the telescope scans along a sky circle
with the polarimeter axis always oriented to
a fixed direction with respect to the radial direction
of the sky circle.  
The observational strategy for $B$-mode measurements
can further be refined by choosing appropriate beam sizes
to suppress the accidental $E$-mode leakage.  
We present the expected variances 
of the measured $E$ and $B$-mode surface-brightness 
fluctuations as functions of the sky annulus radius and beam size.
The measured peak power is expected to be up to $10\%$ of 
the total power contained
in the CMB polarization fluctuations for both $B$ and $E$ modes.  
The temperature-polarization, or $T-E$, correlation can
be measured by the present method as well.
By adopting this strategy, 
the AMiBA experiment is expected to detect, 
in a single sky annulus, a $1\sigma$ $B$-mode signal 
in $340$ hours. 
\end{abstract}



\section{Introduction}
 
The CMB anisotropy provides direct information 
of the oldest relics of 
large-scale structures left from the Big-Bang.
These CMB photons generally contain two kinds of
information at every frequency band, the intensity 
and the polarization. 
Though the CMB photons are originated at the 
last scattering surface near
red-shift $z\sim 1000$, they can be re-processed along the
light paths before reaching us, notably by the gravitational and Doppler 
effects.  These effects dominantly affect the photon intensity, or the 
temperature anisotropy; they can also affect the photon polarization, 
but only to a much lesser degree.
Thus, unlike the temperature anisotropy, it is believed that the CMB 
polarization is a sensitive probe to
the universe at a specific epoch when the photon temperature is about 
$3000$ degrees and the photon optical depth
$\tau$ about $1/2$.
The CMB polarization is generated at this optical depth through the Thomson
scattering of the pre-existing temperature anisotropy by a background 
of homogeneous free electrons\citep{ree68,bon84,pol85}.

Different normal modes in the primordial
fluctuations yield different characteristics in the 
temperature anisotropy,
thereby yielding different types of polarization patterns. 
It has been pointed
out that these different polarization patterns can be divided into two
classes: the electric-type ($E$) pattern and the magnetic-type ($B$)
pattern\citep{kam97,zal97,hu97}.
The primordial density perturbations (scalar modes) generate only the $E$ pattern,
but the $B$ pattern can only be generated by the primordial
gravitational waves (tensor modes) and vortex motion (vector modes).
Since different patterns 
were originated from different physics, the ability to separate the $E$
and $B$ patterns cleanly has become a measure of success for CMB
polarization experiments in a forseeable future.

Following the conventional approach to temperature anisotropy, most past works
that addressed the CMB polarization tended to focus on how the two kinds 
of polarization patterns may be separated in the Fourier domain.  Such an
approach is natural
for the wide-sky maps, such as those planned for the MAP and Planck missions.
However, for the ground-based and ballon-borne experiments, 
the wide-sky maps are impossible to obtain,
and the adopted strategy is often deep exposure of a 
small-patch of sky\citep{sta99,whi99,lo00}. 
Unlike the wide-sky map, the local maps cannot naturally satisfy
the periodic boundary conditions, and hence a brute-force Fourier analysis
to extract the power spectrum
can often introduce distortion to the original data and produces
un-desirable effects. 
The problem is particularly acute for the $B$-mode
measurement due to the small $B$-mode signals.
For this reason, the real-space analysis of local
maps has been called for\citep{sel98,zal98,chi00,ma00,teg00}.

In the past years, an alternative approach has also been examined 
in an attempt to alleviating 
the problem of boundary conditions\citep{zal98,rev00}.  
When the measurements are taken along
a ring on the sky, the data will automatically satisfy the periodic
boundary condition with discete wavenumbers.  As the ring diameter can
be over tens of degrees, the large sky coverage of the ring
permits one to extract a wide spectrum of discrete modes from the 
measurement.  Though this approach solves the problem of boundary
conditions, the separation of $B$ and $E$ modes
is still not natural, as it requires precise
short-term calibration of detectors within the individual ring\cite{ode00} 
and long-term calibration across different rings\cite{rev00}
in order to extract the appropriate Stokes $Q$ and $U$ to construct
$E$ and $B$ modes.

Recently, it has been proposed that with a 
suitable projection of the real-space
data, one is able to separate the $E$ and 
$B$ patterns directly from the local
map without resorting to the Fourier analysis\citep{chi00,ma00}.  
They also illustrate what the visual appearances of the unfamiliar 
$E$ and $B$ tensor patterns are in the projected vector space.
In these previous works, the polarization tensor is first 
projected into a vector and a 
pseudo-vector field through differentiation of the
measured data, and these vector fields are further 
projected into a scalar and a pseudo-scalar through a contour 
integration of {\it arbitrary} shape.  The scalar component
results from the $E$ mode and the pseudo-scalar from the $B$ mode.

Though illuminating, this proposed projection, 
however, suffers from a differential operation on the data for projecting
the polarization tensors into vectors, which invariably amplifies
the measurement noises.  Despite that the differentiation is 
followed by a contour integration to construct the projected
$E$ and $B$ scalars, which damps out the amplified noises,
the noise damping can hardly compensate 
for the noise amplification.  Hence it is desirable to
pursue a different method that can also cleanly separate the $B$ 
and $E$ patterns in the real space.

A tensor (vector) can be projected into a vector (scalar) through a 
vector operation, which can either be a projection-derivative or a
projection-integration along a given contour.  Both 
have been used in our previous proposal\cite{chi00,ma00}.
As the differentiation amplifies the data noise, the former is 
ruled out.  For the latter, various projection integrations are available.
As will be shown in this paper, the only viable ones, useful
for separating the pseudo-tensor patterns from the tensor pattern
are those that assume circular integration contours.  
It has been pointed out that circular-scan observations allow
the $E$ and $B$ modes to be separated\citep{zal98}.  The 
present work reinforces this observation and shows that  
the scanning observation along circles is indeed the only
one strategy that permits the $B$ and $E$ modes to be
cleanly separated without involving data differencing.

This result is particularly relevant for the  
scanning observations conventionally conducted by, for example, 
balloon-borne and ground-based experiments. 
The recommended circular scanning
observation is nothing but integration of data along a given sky 
circle over a long period of time without distinguishing 
variations on the circle.  This is to be
contrasted to integration along a sky contour of arbitrary 
shape proposed 
by \citep{chi00}, and contrasted to extraction of discrete modes on a sky
circle analyzed by \cite{zal98}.  
In fact, to achieve the goal of separating $E$ and $B$ modes,
it is also necessary that the polarimeter axis of the detector
also rotates as the telescope sweeps along the sky circle. 
Be precise, the polarimeter axis must always orients toward a fixed 
direction with 
respect to the local radial direction of the sky circle.
The AMiBA telescope can have this capability\citep{lo00}. 
We will assess how AMiBA
is expected to perform using this observational strategy.

This paper is organized
as follows.  Sec.(2) gives the formulation, with which the
constraint on the shape of the integration contour is derived.  The 
implications for the projection integrals with circular contours 
are examined in Sec.(3).  Sec.(4) addresses
the measurement of the $E$ modes using this method.
Sec.(5) discusses the measurement strategy for the much weaker $B$ modes.  
Correlation of $E$-mode polarization with temperature anisotropy
is addressed in Sec.(6).  In Sec.(7),
we study how a finite beam width can affect the sharp-beam cases
considered in the previous sections, and show that the ability of the
circular-scan strategy for separating $B$ modes from $E$ modes can
be much improved via a finite-size beam.
Discussions on the AMiBA performance are given in Sec.(8).
Conclusions are given in Sec.(9).
Throughout this paper, we adopt the flat-sky approximation since the CMB
fluctuations of interest are at most few degrees in size.  

\section{Projection Contour Integral}

The polarization tensor ${\bf P}$ can be expressed as\citep{chi00}
\begin{eqnarray}
{\bf P}&&=Q\sigma_x+U\sigma_y =
[\nabla\nabla-(\hat z\times\nabla)(\hat z\times\nabla)] f
+[\nabla(\hat z\times\nabla)+(\hat z\times\nabla)\nabla] g \nonumber\\
&&=Q+iU=4\frac{\partial^2}{\partial\bar w^2}(f+ig)
\end{eqnarray}
where $\sigma_x$ and $\sigma_y$ are the "3" and "1" components of the Pauli
matrices, respectively, and $f$ and $g$ are real functions denoting the
$E$ and $B$ components,
respectively.   Moreover, from the third equality on, we have used a
complex representation for ${\bf P}$, where
$w=x+iy$ and $\bar w=x-iy$.
A convenient way to construct a scalar from ${\bf P}$ proceeds as follows.
Pick a closed contour $C$ on the polarization field,
contract the tensor ${\bf P}$ with a suitable local tensor to form a
scalar/pseudo-scalar at every point on the contour, and then
sum up these scalar quantities over
the contour to obtain a contour-integrated value.
Mathematically, the operation can be written as
\begin{equation}
Y\equiv 4\int_C\frac{\partial^2 (f+ig)}{\partial\bar w^2}\bar wd\bar w.
\end{equation}
where $\bar w d\bar w$ is the local tensor for projecting out the scalar $Y$,
$C$ is the closed contour along which the line integration is performed, and the
coordinate $\bar w$ can be chosen so that the origin is inside the contour.
In terms of the Stokes $Q$ and $U$, Eq.(2) is
\begin{equation}
Y=\int_C[(xQ+yU)dx-(yQ-xU)dy]-i\int_C[(yQ-xU)dx+(xQ+yU)dy].
\end{equation}

Considering the fact that $\bar w\partial^2 A/\partial\bar w^2$ can be expressed
as a total derivative, equal to
$\partial[\bar w(\partial A/\partial\bar w)-A]/\partial\bar w$, and that
$f$ and $g$ are both analytical functions, we arrive at the following 
expression for
$Y$ of Eq.(2),
\begin{equation}
Y=i\int_C[\frac{\partial F}{\partial y}dx-\frac{\partial F}{\partial x}dy],
\end{equation}
where $F\equiv\bar w(\partial(f+ig)/\partial\bar w)-(f+ig)$.
Using the Stokes theorem and the fact that 
$\nabla^2=4\partial^2/\partial\bar w\partial w$, Eq.(4) can further be 
rewritten as
\begin{equation}
Y=-i\int \nabla^2 F d^2 S
=-i\int\bar w\frac{\partial}{\partial\bar w}\nabla^2(f+ig) d^2 S,
\end{equation}
where the surface integration is confined within the area bounded 
by the contour $C$.
Instead of
placing an integration limit for the surface integral, we may multiply the integrand
of Eq.(5) by a top-hat filter function $W(x,y)$ and carry out the surface integral
without an explicit bound, where $W(x,y)=1$ inside the contour $C$ and 
$W(x,y)=0$
otherwise.  This permits a convenient way to evaluate the second equality
of Eq.(5).  Note that 
$\bar w\partial/\partial\bar w=(\partial/\partial\ln r)+i(\partial/\partial\phi)$ 
in the polar coordinate.   With an integration by part, Eq.(5) becomes
\begin{equation}
Y=i\int[\frac{\partial W}{\partial\ln r} + 2W + i\frac{\partial W}{\partial\phi}]
\nabla^2(f+ig) rdr d\phi.
\end{equation}
Note that $W$ is a real quantity, and both real and imaginary parts of 
$Y$ generally contain a mixture
of the real $E$-mode $f$ and real $B$-mode $g$.  However, at some special
arrangements, 
$E$ and $B$ modes can become 
separated, with $g$ contributing only to the real part and $f$ only to the
imaginary part, or the opposite.
These situations only happen when $r^2 W$ depends either on $r$ or on $\phi$,
but not on both.  The choice of depending only on $\phi$
can be ruled out since such a window function is 
not compact.  Thus, the one and only 
one choice for the contour must be such that 
$W(x,y)=W(r)$, and moreover $W(r)=\Theta(R-r)$, a circular top-hat filter
function of radius $R$.  

That is, the only scanning trajectory that is capable of 
projecting out the desired scalar and pseudo-scalar is 
when it is a circle.
The circular scan yields the real part
of $Y$ to be entirely contributed
by the $B$ mode and the imaginary part of $Y$ by the $E$ mode.
It also leaves $Y$ to depend on a single
parameter, i.e., the radius $R$ of the circle.  The above derivation
also proves that a special case of circular-scan observations analyzed
by Zaldarriaga \cite{zal98} is indeed a unique one.  

\section{$E$-$B$ Amplitude through a Circular Filter}

We now examine what quantities of the polarization tensor ${\bf P}$
are to be extracted from the projection integration along a circle.  
It will also be shown that the projection of ${\bf P}$ defined
in Eq.(2) corresponds to physical rotation of the polarimeter axes, so 
that they remain fixed with respect to the local radial direction along 
the circle. 

Let the circular contour of radius $R$ center 
at ${\bf x}_0$, so that $r\equiv |{\bf x}-{\bf x}_0|$.  
It then follows from Eqs.(3) that
\begin{eqnarray}
Y(R;{\bf x}_0)
&&= -iR^2\int_0^{2\pi}[(\cos(2\phi)Q+\sin(2\phi)U)
+i(\cos(2\phi)U-\sin(2\phi)Q)]d\phi\nonumber\\
&&=-iR^2\int_0^{2\pi}
[r\frac{\partial}{\partial r}(\frac{1}{r}\frac{\partial}{\partial r})
-\frac{1}{r^2}\frac{\partial^2}{\partial\phi^2}
+\frac{2i}{r}\frac{\partial}{\partial\phi}
(\frac{\partial}{\partial r}-\frac{1}{2r})](f+ig) d\phi\nonumber\\
&&= i\int_0^{2\pi}\int_0^\infty \bar W(r)
\nabla^2(f({\bf x})+ig({\bf x})) rdr d\phi, 
\end{eqnarray}
where the second equality can be straightforwardly 
derived from Eq.(2) and the third equality is simply
Eq.(6) with a filter function 
$\bar W\equiv (1/r)d(r^2 W)/dr$.
Similar expressions for Eq.(7) have also been derived
before\citep{sel98,cri00}
based on a related idea in weak lensing studies\citep{kai94}.

The angular factors on the right of the first equality
in Eq.(7) are the familiar ones.
It describes how 
$Q$ and $U$ rotate as the second-rank tensor components
when the coordinate rotates.   In fact, these angular factors mean
that the polarization tensor $Q+iU$ must be rotated to 
$Q_0+iU_0$ of a
preferred coordinate $(x_0,y_0)$, where the $x_0$-direction is 
along the local radial direction of the 
sky circle and the $y_0$-direction
along the local azimuthal direction.
This point is central to the present work, in that not only the
scanning trajectory should be a circle, but the polarimeter
axes of the telescope must rotate as well.

The separation of $E$ and $B$
modes depends critically on a perfect
angular integration and perfect polarimeter-axis alignment.
The quantity $Y$ thus measures the angular average of
$4r^2(\partial^2 f/\partial (r^2)^2)$ of the $E$ mode
and $4r^2(\partial^2 g/\partial (r^2)^2)$ of the $B$ mode.
It turns out that these unfamiliar derivatives of $f$ and
$g$ have the same power spectra as those of $\nabla^2 f$
and $\nabla^2 g$, respectively.
The respective power spectra for $\nabla^2 f$ and $\nabla^2 g$
are $k^4|f_k|^2$ and $k^4|g_k|^2$, where $f_k$ and $g_k$ 
are the Fourier components 
of $f$ and $g$.  
Each power spectrum equals the corresponding
power spectrum of the polarization tensor: 
\begin{equation}
\langle|Q_k^{E,B}|^2\rangle+\langle|U_k^{E,B}|^2\rangle=k^4|h_k|^2,
\end{equation}
where $Q_k^{E,B}$ and $U_k^{E,B}$ are the $E$ and $B$ contributions to 
the Fourier components of 
$Q(\equiv(\partial^2/\partial x^2-
\partial^2/\partial y^2)f-2(\partial^2/\partial x\partial y)g)$ and 
$U(\equiv (\partial^2/\partial x^2-\partial^2/\partial y^2)g
+2(\partial^2/\partial x\partial y) f)$, 
respectively, and $h$ can be either $f$ or
$g$.  Equation (8) is very useful since the measured
quantity $Y_{E,B}$ can be related to
the $E$ and $B$ mode amplitudes directly.  In fact,
$\nabla^2 f$ and $\nabla^2 g$ are the topological charge and pseudo-charge
in the polarization tensor field ${\bf P}$\cite{chi00}.

Moreover, $Y_{E,B}(R;{\bf x}_0)$ is a random variable since
the circle may be centered
at any arbitrary origins ${\bf x}_0$. 
The variance of $Y_{E,B}$ can thus be expressed as:
\begin{equation}
(2\pi R^2)\Delta T_{E,B}(R)
\equiv\langle Y_{E,B}^2(R;{\bf x}_0)\rangle_{{\bf x}_0}^{1/2}
=[\int_0^\infty |\bar W_k|^2\frac{k^4|h_k|^2}{2\pi A} k dk]^{1/2},
\end{equation} 
where $\bar W_k$ is the Fourier component 
of $\bar W$, $\langle ...\rangle_{{\bf x}_0}$
stands for the ensemble average of various ${\bf x}_0$, and $A$ for
the area over which the ${\bf x}_0$ surface average is conducted.
Here, $\Delta T_{E,B}$
denotes the measured variance of the 
polarization brightness temperature for mode $(E,B)$, and
the factor $2\pi R^2$ in Eq.(9)
arises from the product of two factors:
a factor $R$, which has been
multiplied to the data to construct
$Y$ in Eq.(2), and the 
circumference $2\pi R$ of the sky circle for the contour integration.
The two factors need to be divided away to obtain the sky surface brightness
$\Delta T_{E,B}$. The measured $\Delta T_{E,B}(R)$ contains the
polarization spectral information of 
either mode, subject only to the
smearing by a $k$-space filter $|\bar W_k|^2$. 
The length scale of smearing depends on the radius $R$, 
but the details of smearing depend on the power
spectrum of the filter function $|\bar W_k|^2$.

Finally, we wish to stress that the separation of $E$ and $B$-modes can, 
in principle, be
achieved with a single-polarization detector.  This is because the 
polarization
axis is always rotated to the appropriate direction at every point 
of the sky circle.  On may, for example, align the polarimeter axis to 
the local radial direction to pick up $E_r$ and integrate
$E_r^2$ over a rotation, and then rotate the axis by $90^o$ to integrate
$E_\phi^2$ over another rotation.
The $E$ mode
signal is simply $\langle E_r^2\rangle-\langle E_\phi^2\rangle$.
For $B$-mode measurements, one may orient the polarization axis to
either $45^o$ or $-45^o$ off the local radial direction, and the
$B$-mode signal is 
$\langle E_{45^o}^2\rangle-\langle E_{-45^o}^2\rangle$.
However, detection of the much weaker $B$ modes with 
single-polarization detectors
may, in practice, suffer from inaccurate calibration of
the two measurements to be subtracted from each other. 

\section{The $E$-Mode Measurement}

Since $\bar W$ consists of a top-hat disk and a thin 
shell, its power spectrum oscillates
in the $k$ space at a particular phase with a 
well-defined frequency comparable to
$2R$, the Gibb's phenomenon, and the 
oscillation amplitude declines as $k^{-1}$.
At the first glance, it appears that such a
$k$-space filter is too extended to
be useful.  However, this disadvantage
may be turned into an advantage for
the polarization measurement. 
The power spectrum of 
$E$ mode also oscillates in the $k$ space
with well-defined frequency and phase.
This allows for the possibility that one
may adjust the scan radius $R$ 
to control the oscillation frequency in the power
spectrum of $\bar W$ so as to optimize the detection 
of $E$-mode signals.  One
would like to match the frequencies of the two oscillations 
to maximize the overlap of
$|\bar W_k|^2$ and $|h_k|^2$ in Eq.(9).  
However, this does not necessarily
produce the maximal power since the two oscillations 
may be out of phase, giving destructive interference.  
The phase of the $k$-space oscillation 
in the $E$-mode power spectrum is 
such that it always oscillates asymptotically 
as $\sin(2\pi k/S)$ for $k/S\ge 2$, 
where $S$ is the oscillation period; 
this behavior is rather insensitive to
the cosmological parameters, such 
as $\Omega_m$ and $\Omega_\Lambda$.
Most power of $k^2(|Q_k|^2+|U_k|^2)$ in
the $E$-mode is contained in between
$k/S=2$ and $k/S=5$. 
It will turn out that $|\bar W_k|^2$ has
an optimal phase such that it also oscillates
as $\sin(2kR)$.  
By adjusting the scan
radius $R$ so that $R\approx 2\pi/S$, we can
obtain an optimal overlap of
$|\bar W_k|^2$ and $|h_k|^2$, whereby considerable
power of the $E$ polarization is captured.

We now turn to show that $|\bar W_k|^2 \sim \sin(2kR)/k$ 
at a large $kR$.  The two dimensional 
Fourier transformation of $\bar W$ is 
\begin{eqnarray}
\bar W_k&&=\int [2\Theta(R-r)-r\delta(r-R)] e^{ik r \cos(\phi-\phi_k)}
d^2{\bf r}\nonumber\\
&&=\int_0^\infty rdr [2\Theta(R-r)-r\delta(r-R)]
\int_0^{2\pi}d\phi
\sum_n J_n(kr) e^{in(\pi/2+\phi_k-\phi)}\nonumber\\
&&=2\pi\int_0^\infty rdr J_0(kr) 
[2\Theta(R-r)-r\delta(r-R)]\nonumber\\
&&=2\pi R^2[2\frac{J_1(kR)}{kR}-J_0(kR)]
=2\pi R^2 J_2(kR),
\end{eqnarray}
where the first term of the third equality 
arises from the disk and the second
term from the shell.  The two can be combined to become
$2\pi R^2 J_2(kR)$.  For $kR \ge 2\pi$, $J_2(kR)$ 
oscillates as $-\cos(\pi/4-kR)/\sqrt{kR}$.
Thus, $|\bar W_k(kR)|^2$ asymptotically 
oscillates at a frequency $2R$ as $const.+\sin(2kR)/k$.

To illustrate how the filter function $\bar W_k$ captures the oscillating
$E$ mode power spectrum, we show 
$|\bar W_k|^2$ for three suitably chosen $R$'s in Fig.(1a),
(1b) and (1c), and compare them with the $E$-mode power spectra of
the $\Lambda$CDM model with $\Omega_\Lambda=0.7$,
$\Omega_c=0.28$ and $\Omega_b=0.02$, the standard 
flat CDM model with $\Omega_b=0.02$, and 
the open CDM model with $\Omega_c=0.28$ and $\Omega_b=0.02$
respectively.  Instead of using $k$, we have used
the conventional spherical-harmonics mode 
number $l$ as the horizontal axis.  Identify ${\bf r}$ to be the 
angle on a flat sky, and the relation 
$l\approx k$ for a large $l$ follows.
Moreover, the conventional power spectrum in the 
spherical harmonic space $C_{E,l}/2\pi$ is replaced by 
$k^4|f_k|^2/2\pi A$ in Eq.(9).  Also shown in Fig.(1) is
another $|\bar W_k|^2$
for a different choice of $R$, where the
first lobe of $J_2^2(kR)$ is just large enough to enclose the major 
peaks in the $E$-mode power spectrum.  Such
an annulus radius is about a factor of $3 - 4$ 
smaller than the choice of $R$ that matches the $E$-mode oscillation. 
These are the two specific $R$'s that are expected to capture a substantial 
power of the $E$ mode.

Plotted in Fig.(2) are 
the expected $\langle\Delta T_E^2\rangle$ as functions of
the circle radius $R$ in Fig.(2) for the $\Lambda$CDM, 
standard flat-CDM and open-CDM cosmologies, 
where $R$ is expressed in unit of arcminute.
The primary peak of the polarization brightness 
corresponds to the smaller one of the two 
specific $R$'s shown in Fig.(1a,b,c).  It is 
expected because the first lobe of $J_2^2$ captures a great deal 
of the $E$-mode power.  This optimal $R$'s
that yield the peaks in Fig.(2) capture about
$10\%$ of the total $E$-mode power for all three cosmologies.

There exist secondary peaks in Fig.(2), which  
partially blend into the primary peaks.
These secondary peaks are given by the larger $R$ that 
produces the oscillating $|W_k|^2$ to match 
the oscillating pattern in the $E$-mode power spectrum. 
Apart from the contributions from the two peaks,
the E-mode surface brightness is seen to drop 
rapidly toward large and small $R$. The $\Lambda$CDM and
standard flat-CDM cosmologies can hardly be
distinguishable at the primary peak except 
for their different amplitudes. 
However, the two cosmologies are distinguishable by the secondary 
peaks.  By contrast, the two peaks in the open-CDM cosmology are 
rather distinct from those of
the other two cosmologies in both scale and amplitude.

It is also noted that for $R\ge 100$ arcmin, the $E$ mode power 
declines by almost a factor of $10$ from its respective peak values in
the three cosmologies.  This is
the scale on which one expects to detect the $B$ mode.  The $E$ mode
on this scale is still much larger than the $B$ mode and may
create serious contamination for the $B$-mode measurement, an issue to
be detailed next.  In Sec.(7), we will show that a finite beam can cure
this problem.

\section{$B$-Mode Measurements}

Although we have shown the plausibility of the circle projection
method for $E$-mode measurements, a naive application of this method to
$B$-mode measurements may find it difficult to realize, due
primary to the expected ten-times-smaller amplitude in the $B$ mode.
As measurements of $B$ modes offer a much greater science
return than measurements of $E$ modes, intense efforts on the $B$-mode
detection are expected to be made in a foreseeable future.
The main challenge here is to defeat not only the noise but also 
the leakage from the much stronger $E$ modes into $B$ modes.
According to Eq.(7), the small real part of $Y$ (the $B$
mode) is actually the residue of cancellation among the 
much stronger $E$-mode signals along the circle.  
Therefore, the experimental setup for $B$-mode measurements can
be very different from that for the $E$-mode.
For example, a telescope that is
capable of measuring the $B$ mode in a reasonable time span
should have a large number of
detectors to speed up the observation.  In addition, the strategy
using single-polarization detectors to measure $B$ mode will likely
fail, due to the much weaker signals in the presence of stronger 
$E$ modes.  Hence, dual-polarization detectors are needed for constructing
the Stokes $Q$ and $U$ simultaneously.

We shall first briefly review the characteristics of 
$E$ and $B$ modes.
The $E$-mode power per log-waveband (i.e., $\log k$)
peaks at the angular scale about $15 -- 20$ arc-minutes
for the currently most favorable $\Lambda$CDM cosmology, c.f., Fig.(1a).
On the other hand, the $B$-mode power per log-waveband is less
than 1$\%$ of that for the $E$-mode and peaks on the angular scale
of a couple of degrees.   
Adopting the circular projection integration,
we show in Fig.(3 a, b, c) how the best choice of 
$|W_k(R)|^2$ may capture the $B$-mode 
power spectra for the three cosmologies indicated in Fig.(1).    
Plotted
in Fig.(4) are the squared variances $\Delta T_B^2(R)$ given
by Eq.(9).  Note that 
the peak power captures about $10\%$ 
of the total $B$-mode power for all three cosmologies.
It is fortunate that the
$B$-mode power concentrates at a much larger scale than where the $E$-mode
power does by almost one order of magnitude.  The scale separation
can help extract the weak $B$ mode from the CMB
polarization in the presence of the strong $E$ mode by adopting an appropriate
beam size.  This is an important subject to be discussed in Sec.(8).

The $N$ dual-polarization detectors can be regarded as $N$ independent 
telescopes, which observe the same
sky circle so as to enhance the signal collection power.    
When each detector rotates around the sky circle at a constant rate,
the angular integration becomes time integration.  Moreover, each
detector does its own time integration without referring to others, 
and therefore
the laborious gain calibration across $N$ detectors is not needed 
frequently.  
The leakage from $E$-modes to $B$-modes within each detector can 
in principle be
significantly removed after many periods of rotation, a subject to be
discussed 
shortly.  The total signal is obtained by summing
up the signals collected by each detector after many rotations.

The aforementioned scale separation of $E$ and $B$ modes
yields $E$-mode variations at discrete frequencies $n\Omega$,
which is about 
several to ten times higher than the discrete frequencies of $B$ modes, 
where $\Omega$ is the rotational
frequency and $n$ an integer. 
The present projection method is nothing more than to extract
the dc component of the signals.  In principle, all $E$ and $B$ modes
are periodic in time, and hence the ac signals can be exactly removed
in one rotation.  
In practice, there can be errors introduced in this
observing strategy, i.e., the errors in sweeping an exact sky circle
at an exact constant rate $\Omega$.
These are time-dependent drifts associated with mechanical controls 
and tend to be of low frequency.
They introduce a phase error, 
i.e., multiplying $c\exp({i\sigma_\omega\alpha(\phi)})$
to the integrand of Eq.(7), where $c$ is the efficiency in picking
up the $E$-mode, 
$\sigma_\omega$ characterizes the error magnitude and
$\alpha(\phi)$ is some slow function of $\phi$, which 
is not necessarily periodic in $\phi$.  If such an error
is a random error instead of a systematic error, it is equivalent to 
broadening the discrete line frequency, 
$n\Omega\to n\Omega+\Delta\omega$ with a small random $\Delta\omega$ of
variance $\sigma_\omega$.
That is, one should expect that in reality the measured
polarization 
fluctuations are no longer periodic in time, and therefore it may yield
some level of $E$-mode leakage into the $B$ mode.  In the worst case, 
the efficiency of coupling to $E$ modes $c=1$, and we shall take this
limit for the estimation of the leakage below. 

The ac $E$-mode leakage,
$\exp[i(n\Omega+\Delta\omega)t]$ of finite $n$, acquires a complex gain factor
\begin{equation}
G\approx\frac{\sum_{j=0}^N e^{i\Delta\omega(t_j)T}}{n\Omega}
\end{equation}
after an integration time $T=2L\pi/\Omega$, where the integer $L$ is
chosen to satisfy $L\sim \Omega/2\sigma_\omega$.
But the dc $B$-mode signal gains by a 
factor $NT$ during the same time interval.  
The numerator of Eq.(11) can be regarded as
random walks with a unit step size and $N$ represents the number of
steps.  Hence 
$|G|$ increases with the integration time as $N^{1/2}/n\Omega$,
and the signal-to-leakage ratio increases 
with time as $N^{1/2}n\pi\Omega/\sigma_\omega$.  

This estimate is only correct for a finite $n$.  
The leakage contributed by the
$E$ mode of $n=0$, which is nearly static and
modulated by the low-frequency noise, requires a separate estimate.
As mentioned in the last paragraph of Sec.(4), the $E$-mode power, over
a circle of a couple of degrees, 
is reduced by almost a factor of $10$ from its peak value to a level 
$\sim 8\times 10^{-13} K^2$ for the $\Lambda$CDM cosmology.
Note also that the gain factor $G$ is complex,
and it is only the imaginary part that can make 
the dc $E$ mode leak into the dc $B$ mode.  Thus,
the leakage power from the dc $E$ mode is
$\sim 8\times 10^{-13}$K$\times (Im G)^2$ after an integration time $NT$,
where   
\begin{equation}
Im G\sim  T\sum_{j=0}^N\sin(\Delta\omega(t_j) T).
\end{equation}
It follows that $|Im G|\sim TN^{1/2}$ and the
signal-to-leakage ratio
increases with time as $N^{1/2}$, 
which is a factor of $ \sigma_\omega/n\pi\Omega (\ll 1)$ smaller than the
leakage caused by the ac $E$ modes.  

In other words, the dc $E$ mode is the dominant
contributor to the leakage.   From this estimate, we notice that 
the extreme low-frequency sweeping errors,
such as $1/f$ noise,
may seriously hamper the $B$-$E$ separation. 
In the estimate below, it is found that $N$ needs to be
of order a few hundreds to suppress the $E$-mode leakage at a 
satisfactory level.  
Nonetheless, such extreme low-frequency noise 
can be efficiently removed
when the calibration of the scanning trajectory is conducted 
periodically. That is, periodic calibration can
modulate the low-frequency noise with a finite-frequency,
thereby making such noise under control.

To be specific for estimating the number of rotations required
to suppress the $E$-mode leakage, we take the $\Lambda$CDM cosmology
as a fiducial model.  The peak $B$-mode power is 
$3.6\times 10^{-14} K^2$
at the circle radius $R\sim 2$ degrees  (c.f., Fig.(4)).
On the other hand, the expected $E$ leakage power at this ring radius is 
$8\times 10^{-13}N^{-1}K^2$ (c.f., Fig.(2)).  
Hence, the minimum $N$ is about $70$ in order to suppress the leakage down
to the $30\%$ level of the expected signal power, and the total number of
rotations is therefore $70\times (\Omega/2\sigma_\omega)$.
As mentioned earlier, the lower bound of $\sigma_\omega/\Omega$ can be
set by periodical calibration.
Take $\sigma_\omega/\Omega\sim 0.1$ as an example; one needs $350$ 
rotations to beat down the $E$-mode leakage satisfactorily.
For a total
integration time several thousand hours (discussed in Sec.(8)), 
it requires a rotation rate at least one turn per day.
In fact, the leakage from the dc $E$-mode can be further reduced by
the finite beam size.  This is the most effective means
for $E$-mode suppression and will be elaborated in
Sec.(8) after the effects of finite beam width are examined in Sec.(7).

We now turn to brief discussions on the potentially
most serious contaminants to the $B$-mode
measurement.
The contamination of the scattered polarized 
photons from the terrestrial
environments, such as the 3-degree CMB photons 
reflected by the ground and
observational instruments, can be a serious 
problem for the polarization
measurement.  This problem is even more acute for 
the $B$-mode measurement.  Nevertheless,
uniform illumination of polarized photons 
to the detector can
contribute no net polarization signal after 
a perfect line integration as the first
equality of Eq.(7) shows; neither can the 
dipole pattern of illuminating sources.
In fact, among all multipole patterns, 
only the quadrupole patterns
can contribute to the polarization 
measurement for the present observing strategy.
Serious measures must be taken to guard against
the quadrupole-moment scattered lights in the experiment.   

The confusion from polarized point 
sources can also be a serious concern
for the CMB polarization 
measurement, and they generally contribute almost equally
to both $E$ and $B$ modes.  They are often
synchrotron and dust emission sources; the former is strong at low radio
frequencies and the latter dominates at high frequencies\cite{lin02}.  
Since the present method measures small 
patches of sky in the real space, it is
possible to identify the point-source 
suspects through measurements at very low and very high
frequencies {\it a priori} and to avoid these regions.
Point-source confusion may thus be a soluble problem.

\section{Temperature Anisotropy-Polarization E-Mode Correlation}

The $E$-mode polarization in the CMBR results from the Thomson scattering
off the temperature anisotropy of the primordial adiabatic and/or
iso-curvature fluctuations.  Therefore, the polarization $E$-mode is 
causally related to the temperature anisotropy, with its phase locked to 
temperature fluctuations.  The correlation between the temperature anisotropy 
scalar field and the polarization tensor field is conventionally defined in the
Fourier space as $C_{TE,k}=-k^2\langle Re[k^2 f_k T_k^*]\rangle$ 
per $\log(k)$ band, 
where $\langle ...\rangle$ denotes the angular average over the direction of the 
wave-vector ${\bf k}$.  Plotted in Fig.(5,a,b,c) are $C_{TE,k}$ of the
same three cosmologies as those indicated in Fig.(1).  The averaged signals 
are a factor of few stronger than those of the $E$ polarization.
In the real space, the inverse Fourier transformation gives
\begin{equation}
C_{TE}(r)=\int \frac{d^2{\bf x}_0 }{A}
\nabla^2 f({\bf r}+{\bf x}_0)
\Delta T({\bf x}_0).
\end{equation}

The measured Stokes parameters and temperature anisotropy
are subject to different filter functions.  
While the
filter function for the polarization tensor field is $\bar W$, 
the filter function 
$W_s$ for the $\Delta T$ scalar field is
$W_s(r;R)=r\delta(r-R)$.  The Fourier component of $W_s(r;R)$ is 
$W_{s,k}(R)=2\pi R^2 J_0(kR)$.
We define
the $T-E$ correlation by multiplying $Y_E$ on the sky circle
to the temperature anisotropy on the same circle.
The filter-weighted $T-E$ correlation, in unit of temperature, is thus
\begin{eqnarray}
\bar C_{TE}(R)&&=(2\pi R^2)^{-2}\langle Y_E(R;{\bf x}_0)
\int W_s(r;R)\Delta T({\bf x}_0+{\bf r}) d^2{\bf r}
\rangle_{{\bf x}_0}\nonumber\\
&&=-\int_0^\infty k^2\frac{Re[f_k T_k^*]}{2\pi A} J_0(kR) J_2(kR) kdk.
\end{eqnarray}
Plotted in Fig.(5 a, b, c) are also the composite filter function 
$J_0(kR)J_2(kR)$
of some choices of $R$ for the three cosmologies.  Figure (6) shows
$\bar C_{TE}(R)$ as functions of $R$ in unit of arcminute.  As
$C_{TE}(R)$ is not a positive definite quantity, the spatial 
filtering significantly
smears out the $T-E$ correlation and
the resulting $\bar C_{TE}$ gives a relatively
small signal strength for all filter sizes $R$.  Thus, 
applying the circular-contour integration to measure $T-E$
correlation is probably not an optimal strategy. 
Nevertheless, if an experiment 
has a sufficiently high sensitivity 
to detect the $E$-mode variance, it should also be well within its capability
to measure $\bar C_{TE}$.

\section{Effects of Finite Beam Width}

In a real experiment, the detector beam has a finite width.
Even with a finite beam width, the present strategy for separating $E$ 
and $B$ modes still holds.
Below, we specifically consider the scanning-beam observation with a
single detector, where the polarimeter axes are rotated to align with the 
local radial and azimuthal
directions defined by the beam-center position ${\bf R}$.  Moreover,
the beam also co-rotates
with the polarimeter axes so that its orientation remains fixed 
with respect to the local radial direction.  Such a setup measures
the following surface brightness weighed by the beam around the circle:
\begin{eqnarray}
Sb({\bf r})
=&&-i\int_0^{2\pi}\frac{d\phi_R}{2\pi}A({\bf r}-{\bf R})
[(\cos(2\phi_R)Q+\sin(2\phi_R)U)
+i(\cos(2\phi_R)U-\sin(2\phi_R)Q)]\nonumber\\
=&&-i\int_0^{2\pi}
\frac{d\phi_q}{2\pi} e^{i2(\phi_R-\phi)}A({\bf r}-{\bf R})\nonumber\\
&&[r\frac{\partial}{\partial r}(\frac{1}{r}\frac{\partial}{\partial r})
-\frac{1}{r^2}\frac{\partial^2}{\partial\phi^2}
+\frac{2i}{r}\frac{\partial}{\partial\phi}(\frac{\partial}{\partial r}
-\frac{1}{2r})](f+ig),
\end{eqnarray}
where the second equality of Eq.(7) has been used, $A({\bf r}-{\bf R})$
is the rotating beam pattern, and $\phi_R$ the azimuthal angle of
the beam center.  The beam-center angle $\phi_R$, instead of $\phi$, appears
on the right of the first equality because the rotation of polarimeter axes 
always refers to the beam-center position on the circle.
The beam pattern $A$ can generally be expanded in terms of the monopole,
quadrupole, etc..  When the beam pattern co-rotates
with the polarimeter axis, both angles $\phi_R$ and $\phi$ always appear in 
these multipole moments in
the combination, $\phi-\phi_R$.  Hence when the beam-center angle
$\phi_R$ is integrated out, we obtain an additional radial window 
function to account
for the fact that the finite-size beam measures a sky annulus of 
finite width rather 
than an infinitely sharp ring, as illustrated below.  The radial 
window function is
a real quantity, and therefore the separability of $B$ and $E$ 
modes remains intact.

For the purpose of illustration, we let the beam be symmetric Gaussian with 
a half width $d$, though in practice the beam is not necessary a symmetric one, 
\begin{eqnarray}
A({\bf r}-{\bf R})&&=e^{-({\bf r}-{\bf R})^2/2d^2}\nonumber\\
&&=e^{-[r^2+R^2-2rR\cos(\phi-\phi_R)]/2d^2}\nonumber\\
&&=\sum_n I_n(rR/d^2)e^{in(\phi-\phi_R)}e^{-(r^2+R^2)/2d^2},
\end{eqnarray}
where the circle has been taken to have a radius $R$, 
and $I_n$ is the modified Bessel function.  
The effective area of the beam is $2\pi d^2$.
The angular average of $\phi_R$ of Eq.(15) can then be 
carried out without involving $\phi$, 
and it follows that
\begin{equation}
Sb({\bf r};R, d)
=-i I_2(rR/d^2)e^{-(r^2+R^2)/2d^2}
[r\frac{\partial}{\partial r}(\frac{1}{r}\frac{\partial}{\partial r})
-\frac{1}{r^2}\frac{\partial^2}{\partial\phi^2}
+\frac{2i}{r}\frac{\partial}{\partial\phi}(\frac{\partial}{\partial r}
-\frac{1}{2r})](f+ig).
\end{equation}

In the case when the beam is asymmetric, this
method also works, as long as the beam shape remains fixed with respect to
the local rotating coordinate around the circle.  
For example, when the beam has a
quadrupole component, the beam
$A(\Delta{\bf r}^2/2d^2)(\equiv A({\bf r}-{\bf R})^2/2d^2)$ in Eq.(16)
will need to be replaced by 
\begin{equation}
A[{1\over 4}(b^{-2}+a^{-2})\Delta{\bf r}^2+{1\over 4}(b^{-2}-a^{-2})
((\Delta x^2-\Delta y^2)\cos(2\phi_R)+2\Delta x\Delta y\sin(2\phi_R))],
\end{equation}
where $a$ and $b$ are the long and short axes, respectively, and the
$\phi_R$-dependence accounts for the co-rotating beam pattern.
The quadrupole moment in the argument of $A$ can be re-arranged to 
become $(1/4)(b^{-2}-a^{-2})[r^2(\cos(2(\phi-\phi_R))-1)+
(\Delta{\bf r})^2]$.  
When such a beam is substituted into Eq.(15), the result is similar to
Eq.(17) except for a different radial function  
in Eq.(17).  This nice result 
allows us to carry out the next manipulation
for projecting out the $E$ and $B$ modes as shown below.
(We shall not dwell on asymmetric 
beams in the following discussions,
but only to point out here that asymmetric beams pose no problem 
for the proposed $E-B$ separation.) 

We now perform an angular average of Eq.(17) over $\phi$, corresponding to
retaining only the dc component of the signals, and find that
\begin{equation}
\bar Sb(r; R,d)=
-i I_2(rR/d^2) e^{-(r^2+R^2)/2d^2}
[r\frac{\partial}{\partial r}(\frac{1}{r}\frac{\partial}{\partial r})
(\langle f\rangle(r)+i\langle g\rangle(r))].
\end{equation}
The real and imaginary parts correspond to the $B$ and $E$ modes,
respectively. 
Identical to what we have found in Sec.(2), 
Eq.(19) can be re-written as a surface convolution
with a window function $\bar W$:
\begin{equation}
\bar Sb(r; R,d)=
i[I_2(rR/d^2)e^{-(r^2+R^2)/2d^2}]
\int \frac{d^2{\bf r}'}{2\pi r^2} 
\bar W(r-r')\nabla'^2(f({\bf r}')+i g({\bf r}'))
\end{equation}
where $\bar W(r-r')\equiv 2\Theta(r-r')- r\delta(r'-r)$.
The squared bracket in Eq.(20) sharply peaks at $r=R$ when 
the primary beam width $d\to 0$; it becomes $\delta(r-R)/2\pi R$ and
Eq.(20) is then reduced to Eq.(7) of the sharp-beam case.

Moreover, 
the convolution surface integral of ${\bf r}'$ in Eq.(20) 
can further be expressed as
\begin{equation}
\int\frac{d^2{\bf r}'}{2\pi r^2}\bar W(r-r')\nabla'^2 h({\bf r}')=
\int\frac{d^2{\bf k}}{4\pi^2}J_2(kr)k^2 h_k,
\end{equation}
after an identical algebra as before.  Finally,
the averaged surface brightness over the whole beam becomes
\begin{eqnarray}
\Delta T(R, d)&&=\int d^2{\bf r}\frac{\bar Sb(r;R,d)}{2\pi d^2}\nonumber\\
&&= i\int\frac{d^2{\bf k}}{4\pi^2}
[\int\frac{d^2{\bf r}}
{2\pi d^2}J_2(kr)I_2(rR/d^2)e^{-(r^2+R^2)/2d^2}]
k^2 (f_k+ig_k)\nonumber\\
&&=i\int\frac{d^2{\bf k}}{4\pi^2}J_2(kR) e^{-(k^2d^2/2)}
k^2 (f_k+ig_k)  
\end{eqnarray} 
and its variance
\begin{equation}
\Delta T_{E,B}(R, d)= 
[\int_0^\infty J_2^2(kR) e^{-k^2 d^2}
\frac{k^4|h_k|^2}{2\pi A}kdk]^{1/2},
\end{equation}
where $|h_k|^2$ represents either $|f_k|^2$ or $|g_k|^2$.
When $kd << 1$, we recover the sharp-beam limit, Eq.(9), from Eq.(23). 
Figure (7) plots 
the squared variance of the beam weighed 
surface brightness of $E$ and $B$ modes as a function of
the annulus radius $R$ with various $d$'s for the fiducial
$\Lambda$CDM cosmology ($\Omega_\lambda=0.7$, $\Omega_{cdm}=0.28$ and
$\Omega_b=0.02$).
The non-zero $d$ leads to smearing of the peaks appearing 
in the sharp-beam limit.  Moreover, the drastic decline in the $E$-mode
power for a non-negligible $kd$ is also evident in Fig.(7).  This is
due to the Gaussian cutoff in Eq.(23).  

Similarly, when observed with a finite beam, the $T-E$ correlation,
Eq.(14), becomes
\begin{equation}
\bar C_{TE}(R,d)=
-\int_0^\infty k^2\frac{Re[f_k T^*_k]}{2\pi A}
J_0(kR)J_2(kR) e^{-k^2d^2} kdk.
\end{equation}
Shown in Fig.(8) are the $T-E$ correlation of the fiducial 
cosmology for various
$d$'s. Apparently, the Gaussian
cutoff is also at work when the beam width is non-negligible.

\section{Discussions}

We return to the discussions on the $E$-mode leakage to the $B$-mode
measurements, especially when the effects of finite beam width are
taken into account.
Fig.(7) clearly shows that when the half-beam width $d\sim 17$ arcmin, 
the peak $B$-mode power
decreases from the sharp-beam case by a factor about $1.3$.  But with
such a wide beam, the
$E$-mode power decreases by a factor of $10$, as compared with the
sharp-beam case, at the radius $R\sim 2$ degrees. 
Hence the $E$-mode power is
only a factor of $3$ greater than that of the $B$-mode power
around $R\sim 2$ degrees.   
This is due precisely to
the scale separation between $E$ and $B$ modes; the wide beam 
yields a small $kd$ for $B$ modes but a 
sizable $kd$ for $E$ modes.
That is, one can always choose an appropriate beam to significantly 
suppress the $E$-modes leakage while not to sacrifice the $B$-modes.  
For the $\Lambda$CDM cosmology, a half-beam of $17$ arcminutes
is near the optimal for $B$-mode detection.

We are now in a position to estimate how likely the $B$ and $E$ modes may be
detected by the circular scan strategy in the near future.  The AMiBA 
experiment\cite{lo00} 
is equipped with two separate sets of $19$ dishes of different sizes, 
$30$cm and $120$cm, 
mounted on a $6$m turntable platform.  It has
$20$-GHz wide-band receivers with dual polarization capability and
operates at $90$ GHz, a frequency window in which the Galactic synchrotron
and dust emissions are both minimal.  The AMiBA has been designed to operate 
in the interferometry mode, which has the advantage of easily removing 
systematic errors.  However, AMiBA can also operate in the
single-dish mode, to which the present observing strategy applies. 
In each single dish, both left and right circular-polarization 
detectors collect independent signals, and the two sets of signals are
correlated to obtain the Stokes $Q$ and $U$.  Much like the
conventional interferometric 
measurements, correlation of the two independent signals can
significantly remove the
systematic errors as well as the electronic $1/f$ noise.
With the system temperature $T_s=70$K, bandwidth $\Delta\nu=20$Ghz,
and antenna efficiency $\eta_a=70\%$ for AMiBA, 
the noise level per antenna per polarization after an integration time
$t_{int}$ is $1.22\times(\sqrt{2}T_s/\sqrt{\eta_a^2 t_{int}\Delta\nu})$ 
and it amounts to $1.2$mK$/\sqrt{t_{sec}}$, 
where the factor $1.22$ corrects for the non-Rayleigh-Jeans limit
for $90$ Ghz and $t_{sec}$
is $t_{int}$ in unit of seconds.
With $19$ antennae and dual polarization, the system noise is
reduced by $\sqrt{38}$ to become $200\mu$K$/\sqrt{t_{sec}}$.

Thus, in $3.8$-hour integration, one obtains a noise level 
down to $1.7\mu$K,
which is the expected maximum surface brightness of the $E$ mode with
the $1.2$m AMiBA dish, or $d=4.2'$ in Fig.(7).   In $35$ hours,
a $3\sigma$ $E$-mode signal is expected in a single sky annulus
of radius in between $15$ and $30$ arcminutes.  
The $B$-mode peak strength for the AMiBA $0.3$m dish, 
($d=17'$ in Fig.(7)) is about $0.18\mu$K. 
It therefore requires about $340$ hours to detect an $1\sigma$
signal, or $3000$ hours for a $3\sigma$ $B$-mode
signal in a single annulus of radius $2$ degrees or so.

So far, we have estimated the signals for the single-field measurement,
which are in themselves random fluctuations. 
To beat down the sample variance and construct the $B$-mode curves shown 
in Fig.(7), one needs a more efficient approach
for determining the variance of the random signals in the presence
of the instrument noise than the deep measurement described above.
The combined error
$\delta(\Delta T^2)$ of
the sample variance and instrument noise is estimated to be
\begin{equation}
\frac{\delta(\Delta T^2)}{\Delta T^2}
=\frac{\sqrt{2}(1+\sigma_n^2/\Delta T^2)}{\sqrt{N_{sam}}},
\end{equation}
where $\sigma_n$ is the variance of the instrument noise per sky annulus
and $N_{sam}$ the number of independent sky annuli measured.
Note that 
$\sigma_n^2\propto t_{int}$, $\Delta T^2\propto
t_{int}^2$ and the
total observing time is $N_{sam} t_{int}$.  For a fixed total
observing time, we want the single-field
integration time $t_{int}$ such that $(\sigma_n/\Delta T)^2\sim 1$.
That is, the optimal strategy for the determination of the
curves shown in Fig.(7) is to collect signals with
$S/N=1$ per sky annulus.  Thus, the desired
integration time per annulus for the $B$ mode is therefore $t_{int}=340$
hours for AMiBA with a $17$ arcmin half-beam.
The detemination of $\langle\Delta T_B^2\rangle$ 
at the $3\sigma$ level (including the sample variance)
requires $N_{sam}=72$ or a total integration time $2.5\times 10^4$ hours.
This amount of total observing time has become too long to be realistic.
 
Having discussed the capability of AMiBA for the CMB polarization measurement,
we briefly turn to a discussion on the optimal instrument that may be able to
realistically measure a substantial part of the $B$-mode curve shown in
Fig.(7).
The guiding principle for an optimal telescope is to collect as many photons 
as possible, and it thus requires as large a light 
collecting area as possible for a fixed beam size.  However,
the conventional radio telescopes 
mostly work in the diffraction-beam regime, where $A_{phy}\Omega_b/\lambda^2
\sim 1$ with $A_{phy}$, $\Omega_b$ and $\lambda$ being the physical 
area of the telescope, 
the beam solid angle and the photon wavelength, respectively, for
the reason that these telescopes are designed
to maximize the angular resolution for
conventional observations.
However, the $B$ modes are intrinsically of large scale (degree scale), and
the measurements of $B$ modes thus require low-resolution telescopes.  From
the discussions given in the beginning of this section, we see that a 
half-beam width
about $17$ arcmin can just be appropriate for the $B$-mode measurement.  
At $\nu=90$GHz, the dish diameter of a conventional radio telescope
that yields a diffraction beam of $17$ arcmin is about $0.3$
meters.  Such a conventional telescope is not 
optimized.  An optimized telescope should adopt multi-mode
feeds, where the antenna beam may be made much wider than the diffraction
beam.  That is, for a fixed beam
width, one would like to increase the dish size as much
as possible.  (Note that the surface brightness signal-to-noise ratio 
increases linearly with the dish area.)
As an example, a $2$-meter dish with a $17$-arcmin half
beam is $45$ times more sensitive than
a $0.3$-meter dish with a $17$-arcmin half beam.
A large-dish telescope hence has a tremendous advantage over an instrument
such as AMiBA, which consists of $19$ dishes of $0.3$-meter diameter with
diffraction beams.  With the same noise temperature per detector, 
such a $2$-meter, single-detector telescope gains a factor 
$10 (\propto$(ratio of dish area)(ratio of detector number)$^{1/2}$)  
in detection sensitivity over AMiBA, or a
factor $10^2$ reduction in the integration time, 
thereby making the $B$-mode power spectrum measurement
a feasible task to conduct.

However, there is a caveat for this muli-mode feed
approach.  The beam pattern and the polarization property can be
difficult to control.  To attain the sensitivity given by the above
example, $45$ modes are needed;   
the multi-mode bolometer currently in operation contains $10$ modes 
or so for the Maxima experiment 
(private conversation with Paul Richards).
Despite the techincal challenges, it is certainly worthwhile to
contemplate the next generation experiments for $B$-mode measurements
along this direction.  
 
\section{Conclusion}

In sum, this work addresses how the CMBR polarization 
$E$ and $B$ modes can be
separated with a local measurement of Stokes parameters 
$Q$ and $U$.  For scanning observations, we show that the only
scanning trajectory capable of 
separating $E$ and $B$ polarization patterns are those that are 
circles.  Moreover, the polarimeter axis also needs to rotate along the
circle at the same rate as well.
By varying the annulus radius $R$, one can measure 
$E$ and $B$ modes of various scales.  In terms of the conventional
power spectrum, this circular scanning observation is equivalent to
adopting some particular $k$-space filters.
We have presented 
the expected surface-brightness of $E$ and $B$ modes as well as the $T-E$
correlation measured through these filters. 

Though such a detection strategy contains 
only the gross information of
the polarization power spectrum, it nevertheless exhibits 
the following good features:

(a) The measured peak power of $E$ and $B$ modes can be about
$10\%$ of the total power contained in the CMB polarization fluctuations.  
The relatively high efficiency makes this strategy attractive for the 
detection of $E$ and $B$ polarization fluctuations in the CMB.

(b) The measurement can be conducted on a turntable platform, 
where all detectors 
scan the same sky circle, with each detector performing its 
own integration of Stokes $Q$ and $U$.  
The noise in the $N$ detectors 
are all independent and the data gathered by each detector can be 
straightforwardly summed together to enhance the signal-to-noise ratio.
This sweeping-beam strategy 
is particularly useful for the weak $B$-mode detection, as
the scan can be on an almost perfect circle, which is needed
to avoid the $E$-mode leakage. 
The sweeping-beam strategy also avoids the need 
for frequent calibration across different detectors, thereby making the
observation simple and stable. 

(c) The present observing strategy is perfectly valid when the
beam has a finite size.  This opens up the possibility that
one may effectively suppress the $E$-mode leakage to the measurements
of $B$ modes, and the possibility that the observation of the weak CMB
polarization signals can be made efficiently 
by using a large telescope with a finite beam, a beam that is much larger 
than the diffraction beam.

(d) Even when the beam is asymmetric the circular scan strategy remains valid, 
provided that the beam shape and orientation remain fixed with respect 
to the local axes along the circle traced by the beam center.

(e) The present method measures, in 
effect, the combination of a circular-top-hat-filtered 
and a circular-ring-filtered
polarization patterns.  In the Fourier 
domain, both power spectra of the combined 
filter and the $E$ mode oscillate.  There exists 
a filter (primarily the ring filter) 
of size $R$ for the 
two oscillations to have the same
frequencies and phases, so that a 
substantial power of $E$ mode is
picked up.  It results in a second peak in the $E$-mode 
surface brightness, which can be used to distinguish the 
$\Lambda$CDM cosmology from the SCDM 
cosmology, as shown in Fig.(2). 

(f) The measurement is in 
real space, and it can be sensitive to the
spatially localized non-Gaussian features, which can be either genuine 
CMB signals or confusion radios sources\cite{lin02}.

Thus, with the above advantages,
the present local observational strategy 
offers an alternative way to detect the
CMBR polarization for the ground-based or 
balloon-borne experiments, where the observation is 
confined within a relatively small patch of sky.  This method
is particularly useful for $B$-mode measurements,
since $B$-mode signals are so weak that deep
exposure is indeed needed.

Interestingly, the present method of circular scan
can also be applied to the interferometric
measurements of CMBR polarization\cite{ma01}, 
an observational
strategy that originally aims to directly extract the Fourier 
power spectrum rather than making the real-space map.
The details of how it can be
achieved will be reported in a separate paper.

After the submission of this work, we noted a relevant work
appearing in the literature\cite{lew02}, 
which addresses the extraction of $B$ modes
in a circular patch of sky.  By suitable linear combinations of
signals obtained from different concentric sky annuli, 
it is possible to extract, apart from the axisymmetric modes,  
the non-axisymmetric $B$ modes.

\acknowledgments

We thank the AMiBA team for useful discussions.  T.C would also like
to thank Jeff Peterson and Paul Richards for discussions on the workings 
of the multi-mode feed bolometer. 
This work is supported in part by the National Science Council of Taiwan under
the grant NSC90-2112-M-002-026.

\clearpage
\figcaption{Fig.(1 a,b,c):
The power spectra of $E$-modes $l(l+1)C_{E,l}/2\pi$
(solid lines) for the (a) $\Lambda$CDM with
$(\Omega_\Lambda, \Omega_{cdm})=(0.7, 0.28)$, (b) standard flat-CDM,
and (c) open CDM cosmologies. 
Superposed on the power spectra are the 
(arbitrarily scaled) $k$-space filter functions 
$|\bar W_k(R)|^2(=J_2^2(kR))$
of two particular annulus radii $R$ (dotted and dashed lines), 
which are optimaized to capture the features in the power spectra.
\label{fig1a,b,c}}

\figcaption{Fig.(2):
The squared variances 
$\Delta T_E^2$ as functions 
of the annulus
radius $R$ for the three cosmologies of Fig.(1).  
$\Delta T_E^2$
is in unit of $K^2$ and $R$ in unit of arcminute.
\label{fig2}}

\figcaption{Fig.(3 a,b,c):
The power spectra of $B$ modes $l(l+1)C_{B,l}/2\pi$
(solid lines) for the 
three cosmologies (a,b,c) of Fig.(1).   Superposed on the power
spectra are the (arbitrarily scaled) $k$-space filter functions 
$|\bar W_k(R)|^2(=J_2^2(kR))$ of 
the optimized $R$'s that capture the power
of the $B$ modes (dotted lines). 
\label{fig3a,b,c}}

\figcaption{Fig.(4):
The squared variances $\Delta T_B^2$ as functions of $R$ for
the three cosmologies.  
The units are the same as in Fig.(2).
\label{fig4}}

\figcaption{Fig.(5 a,b,c):
The $T-E$ correlation spectra $l(l+1)C_{TE,l}/2\pi$
(solid lines) for the 
three cosmologies (a,b,c).   Superposed on the correlation
spectra are the 
$k$-space filter functions $J_0(kR)J_2(kR)$ for 
some choice of $R$'s. 
Hardly can any choice of $R$ yield $k$-space 
filter functions that oscillate in phase with the correlation
spectra.
\label{fig5a,b,c}}

\figcaption{Fig.(6):
The filtered $T-E$ correlation $\bar C_{TE}$ as functions
of $R$ for the three cosmologies.   The units are the same
as Fig.(2).   The signal strengths are seen to be substantially
lower than the peaks in Fig.(5).
\label{fig6}}

\figcaption{Fig.(7):
The finite-beam effects on the detected $E$ and $B$-mode surface
brightness as functions of the radius $R$ of sky annulus and of various
half-beam widths $d$ for the fiducial $\Lambda$CDM cosmology.
The reduction of surface brightness is due to the Gaussian
cutoff of finite beam in the window function as indicated in Eq.(22).
\label{fig7}}

\figcaption{Fig.(8):
The finite-beam effects on the detected $T-E$ correlation as functions
of $R$ and of $d$ for the fiducial $\Lambda$CDM cosmology.
\label{fig8}}

\clearpage
\begin{figure}
\plotone{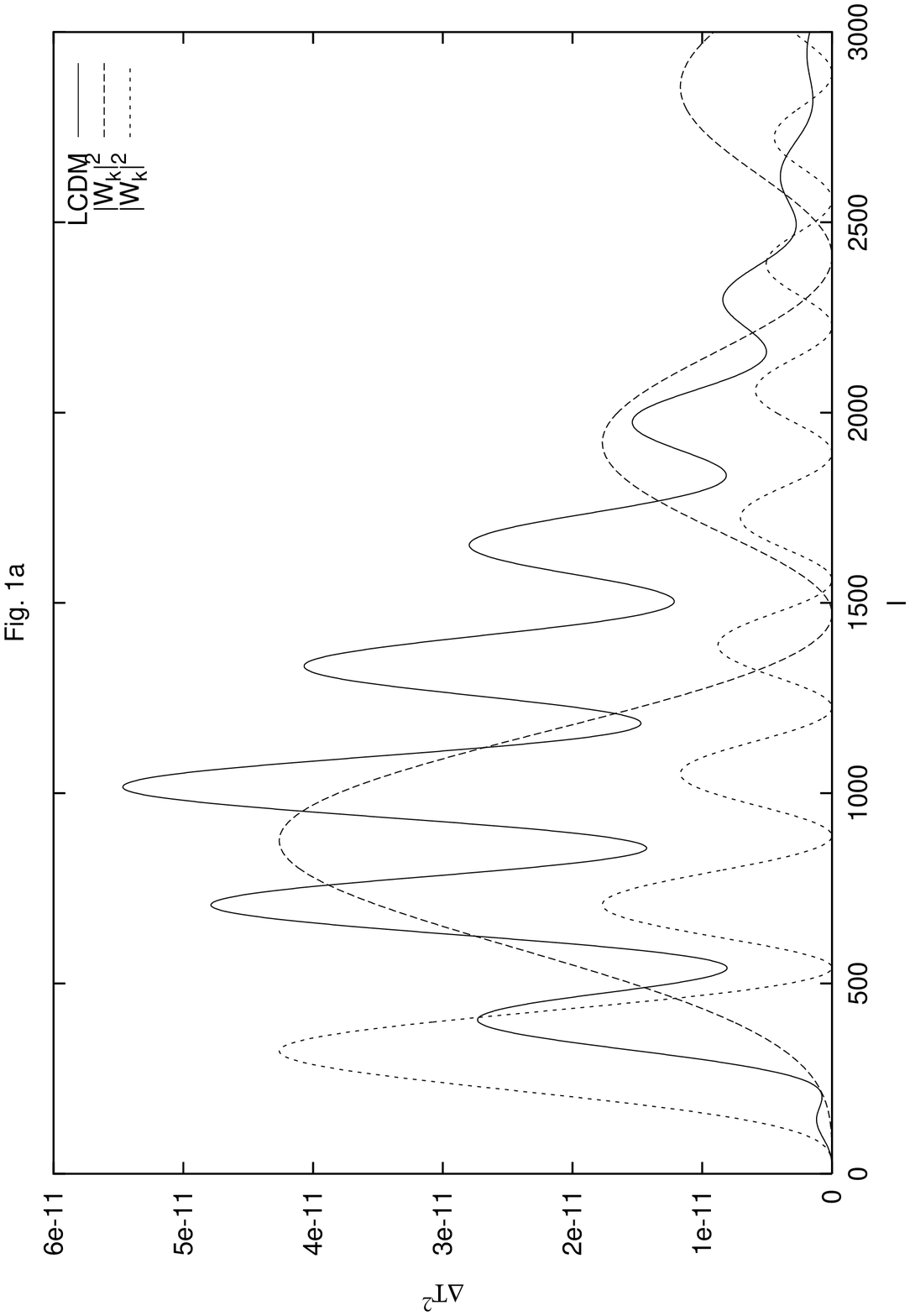}
\caption{
The power spectrum of $E$-modes $l(l+1)C_{E,l}/2\pi$
(solid line) for the $\Lambda$CDM cosmology with
$(\Omega_\Lambda, \Omega_{cdm})=(0.7, 0.28)$. 
Superposed on the power spectrum are the 
(arbitrarily scaled) $k$-space filter functions 
$|\bar W_k(R)|^2(=J_2^2(kR))$
of two particular annulus radii $R$ (dotted and dashed lines), 
which are optimized to capture the features in the power spectra.}
\end{figure}

\clearpage
\begin{figure}
\plotone{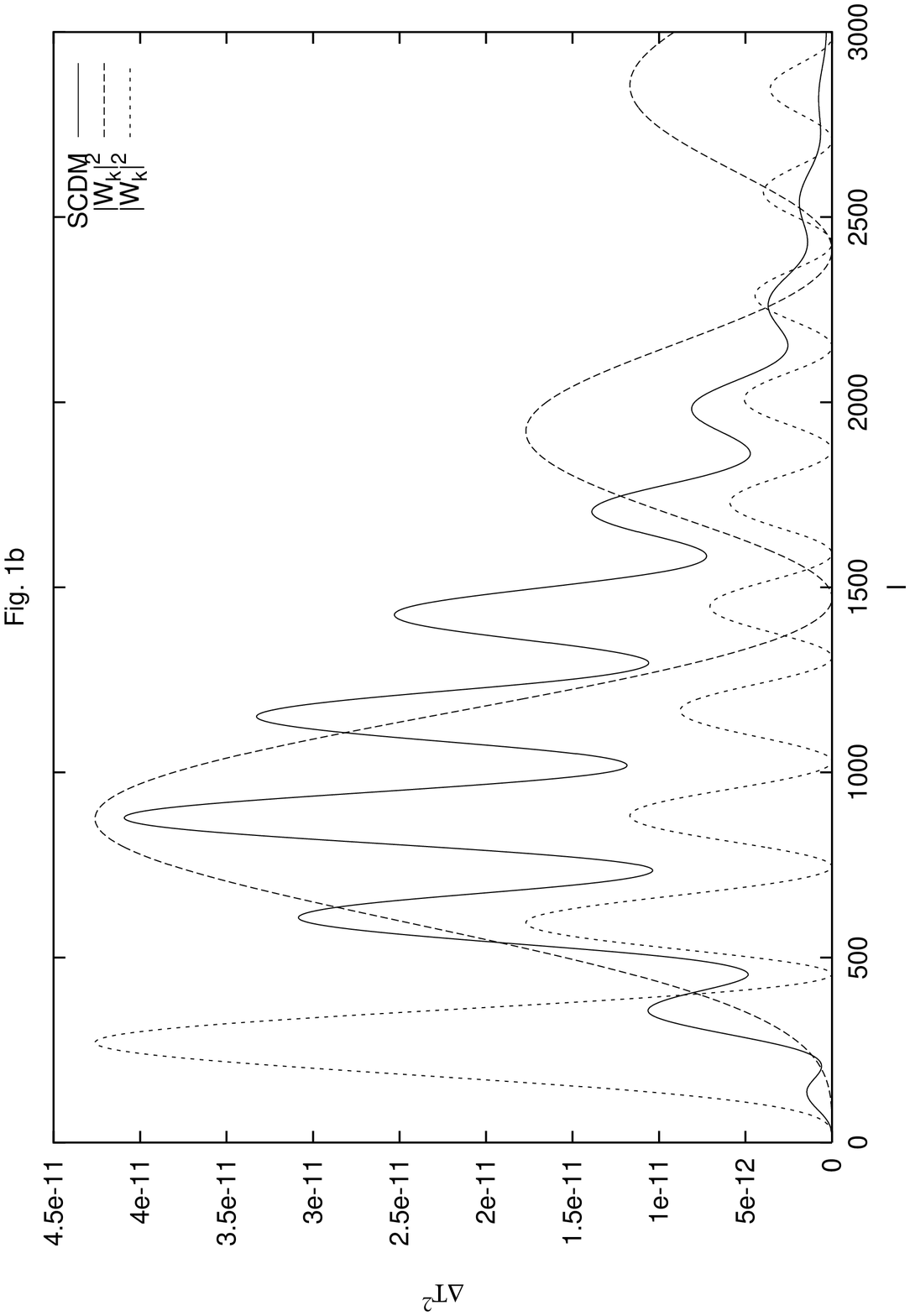}
\caption{Same as Fig.(1a) but for the standard CDM cosmology.}
\end{figure}

\clearpage
\begin{figure}
\plotone{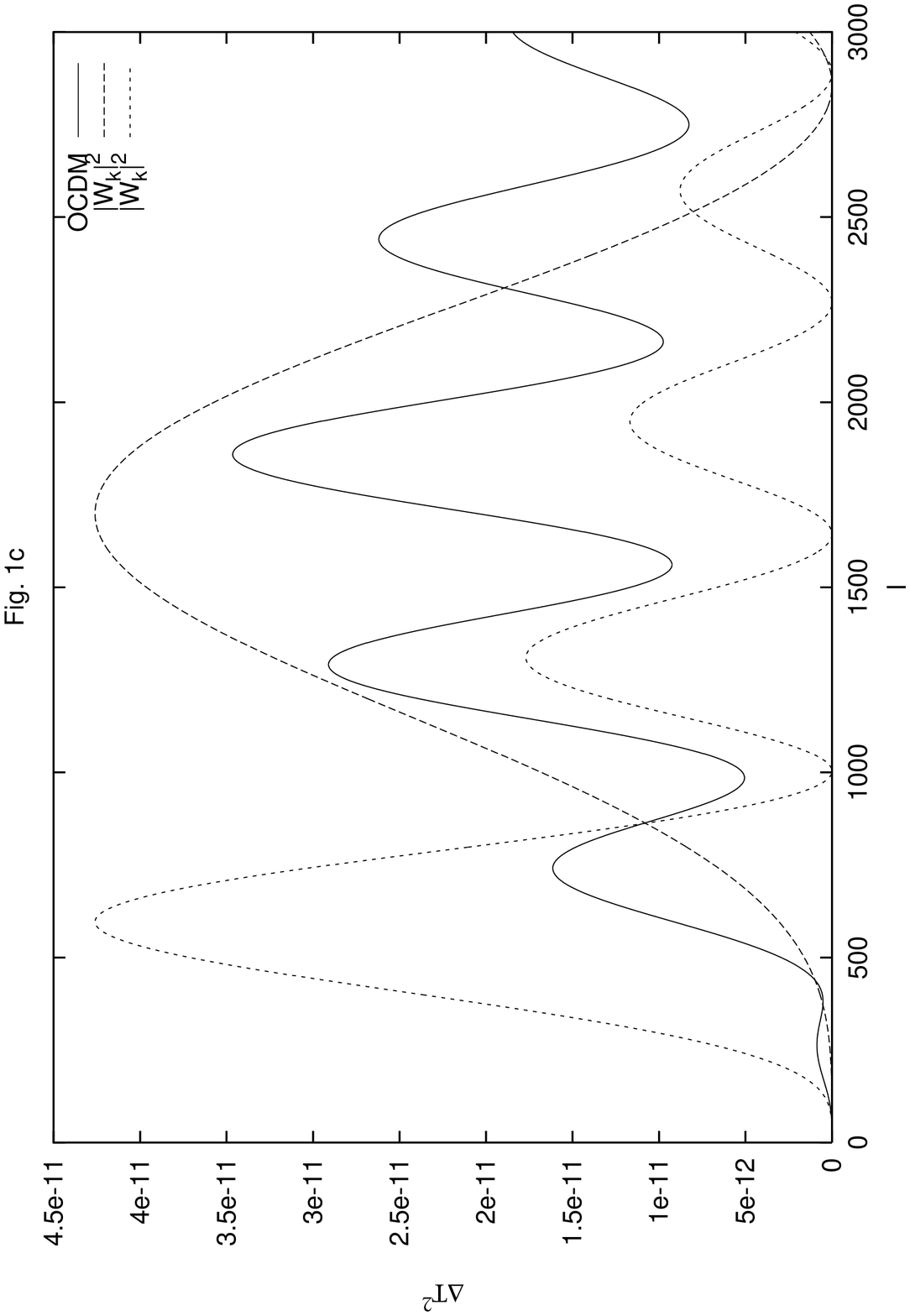}
\caption{Same as Fig.(1a) but for the open CDM cosmology with
$\Omega_{cdm}=0.28$.}
\end{figure}

\clearpage
\begin{figure}
\plotone{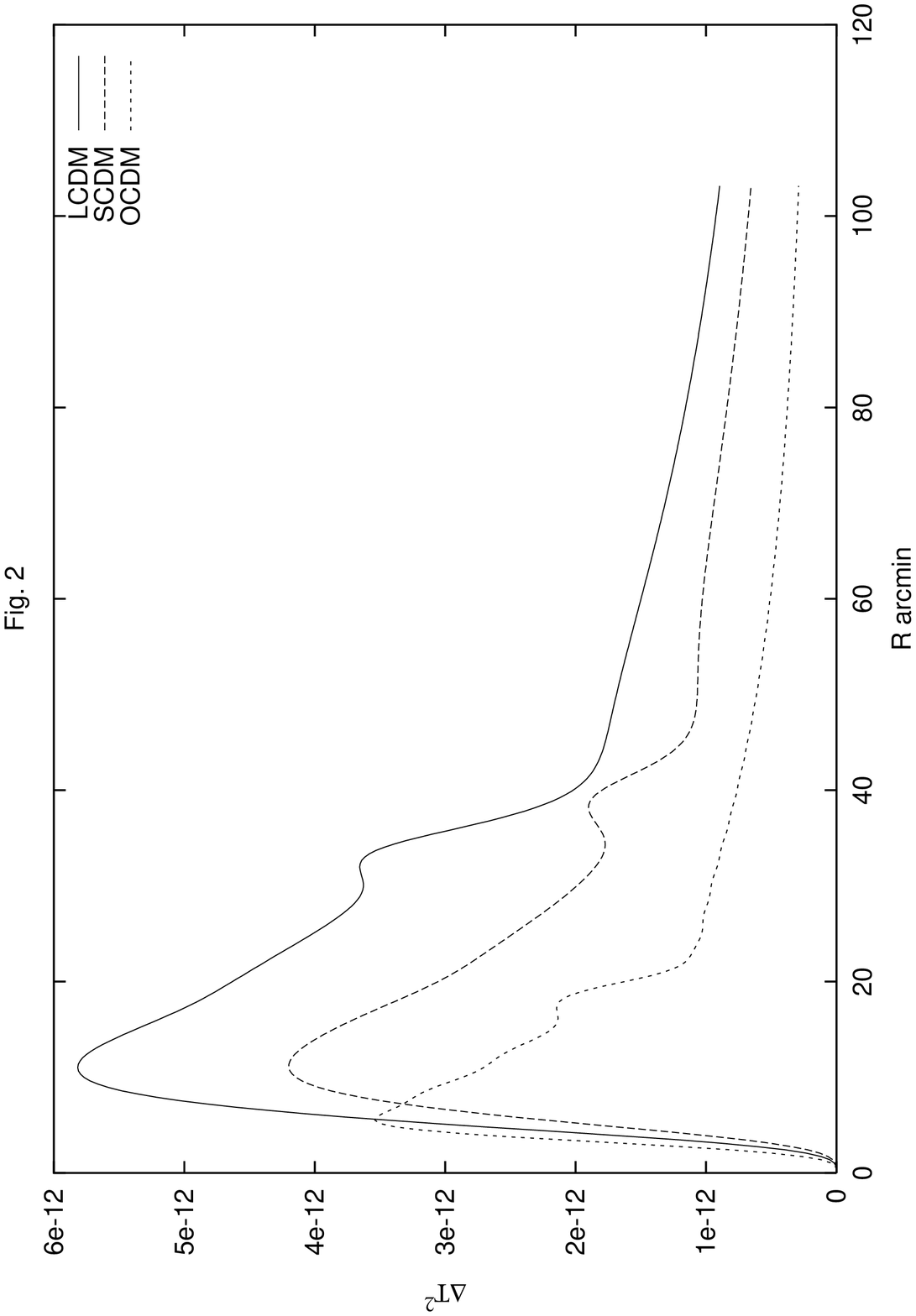} 
\caption{ The squared variances 
$\Delta T_E^2$ as functions 
of the annulus radius $R$ for the three cosmologies indicated in 
Fig.(1 a,b,c).  
$\Delta T_E^2$ is in unit of $K^2$ and $R$ in unit of arcminute.}
\end{figure}

\clearpage
\begin{figure}
\plotone{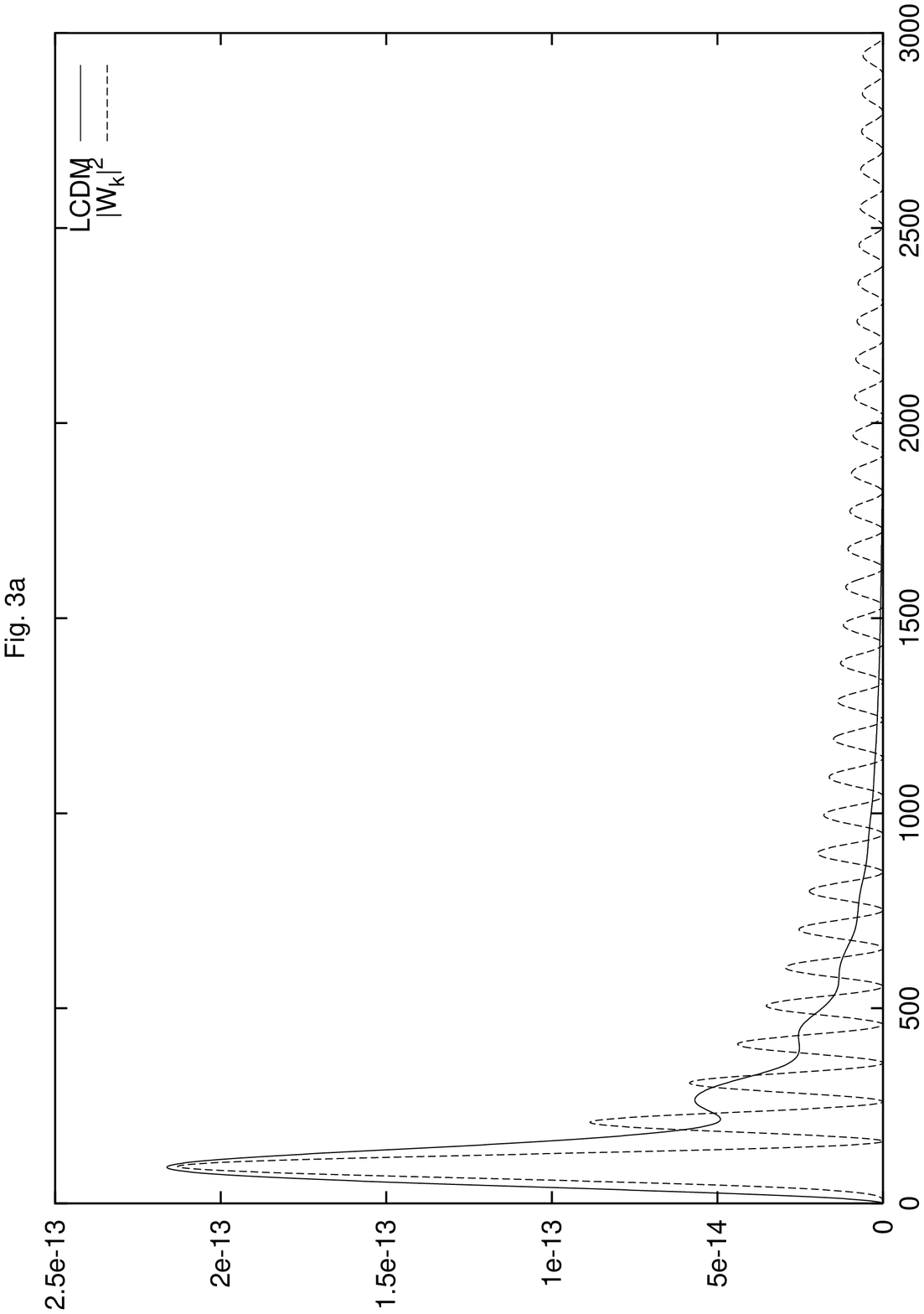} 
\caption{ The power spectrum of $B$ modes $l(l+1)C_{B,l}/2\pi$
(solid line) for the 
$\Lambda$CDM cosmology indicated in Fig.(1 a).   Superposed on the power
spectrum is the (arbitrarily scaled) $k$-space filter function
$|\bar W_k(R)|^2(=J_2^2(kR))$ of 
the optimized $R$ that captures the power
of $B$ mode (dotted line). }
\end{figure}

\clearpage
\begin{figure}
\plotone{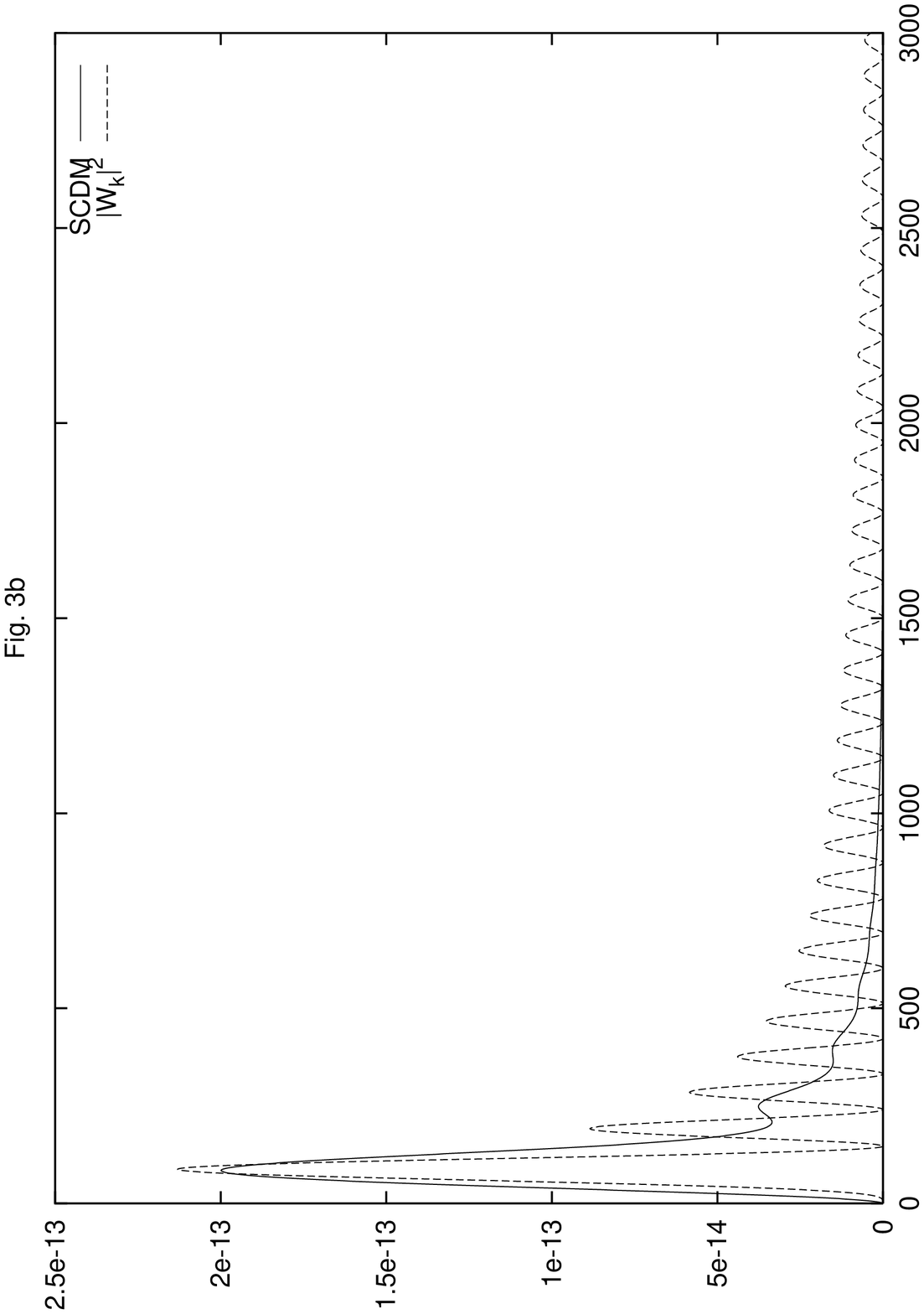} 
\caption{Same as Fig.(3a), but for the standard CDM cosmology}
\end{figure}

\clearpage
\begin{figure}
\plotone{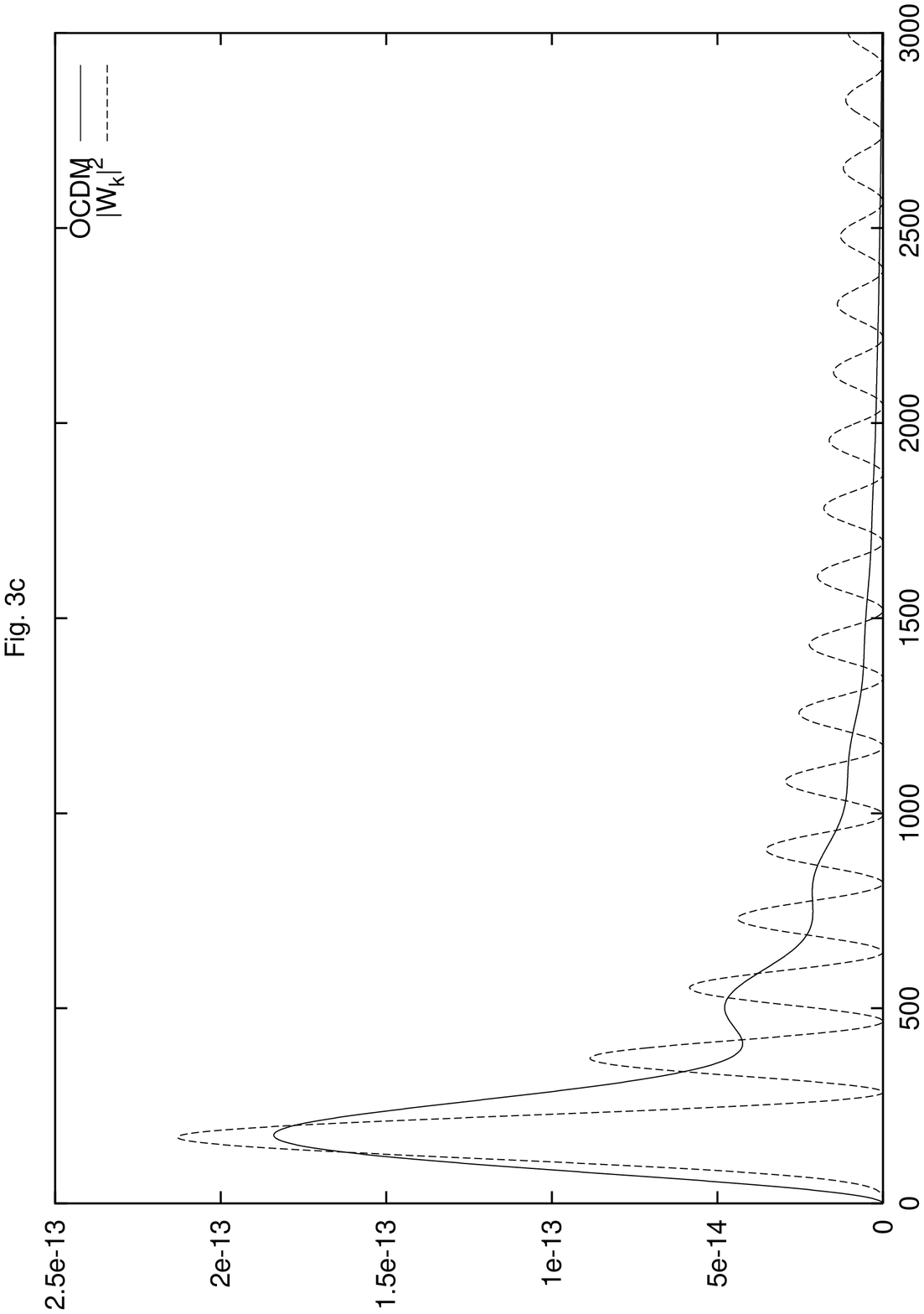} 
\caption{Same as Fig.(3a), but for the open CDM cosmology 
with $\Omega_{cdm}=0.28$.}
\end{figure}

\clearpage
\begin{figure}
\plotone{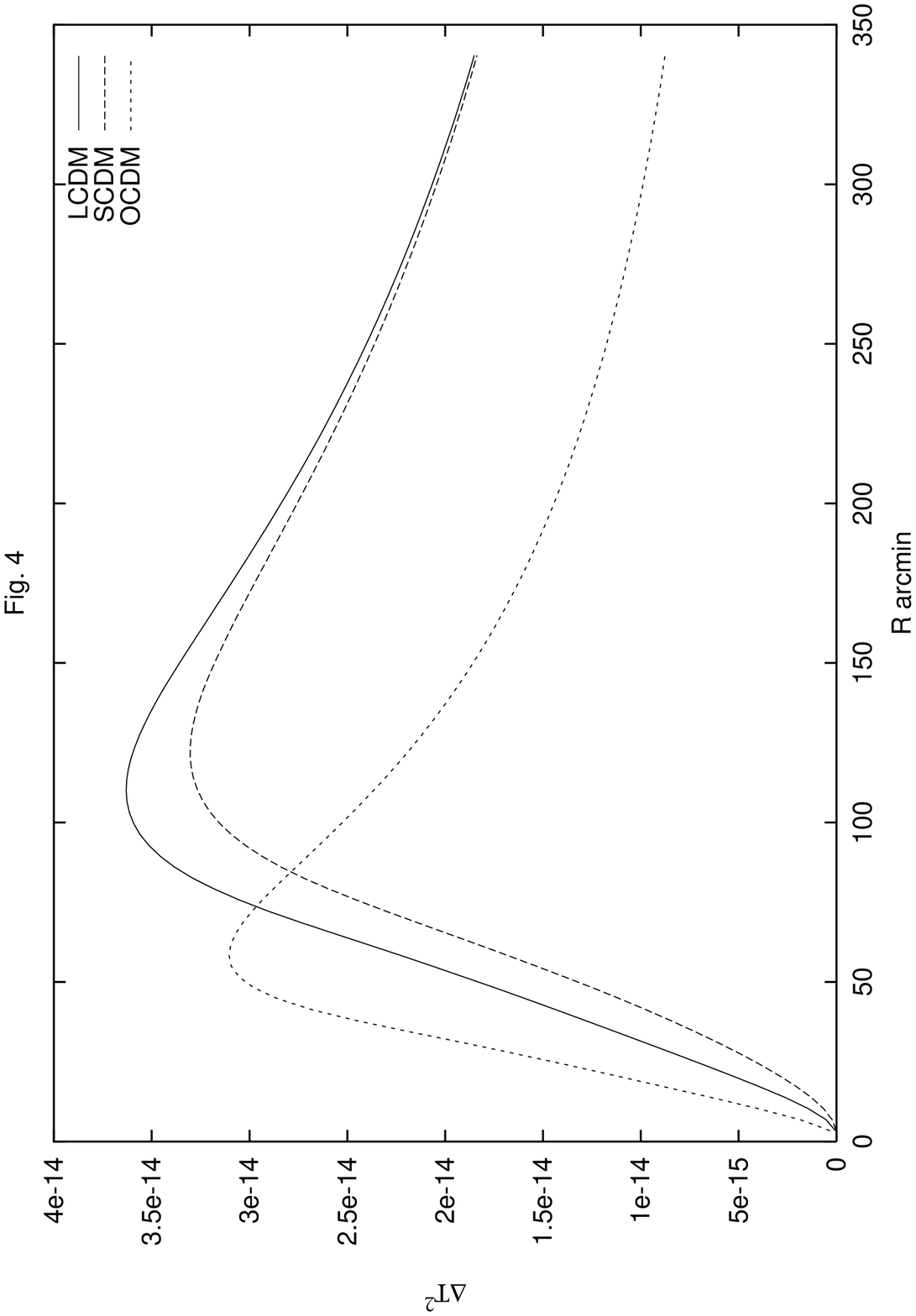} 
\caption{ The squared variances $\Delta T_B^2$ as functions of $R$ for
the three cosmologies.  The units are the same as in Fig.(2).}
\end{figure}

\clearpage
\begin{figure}
\plotone{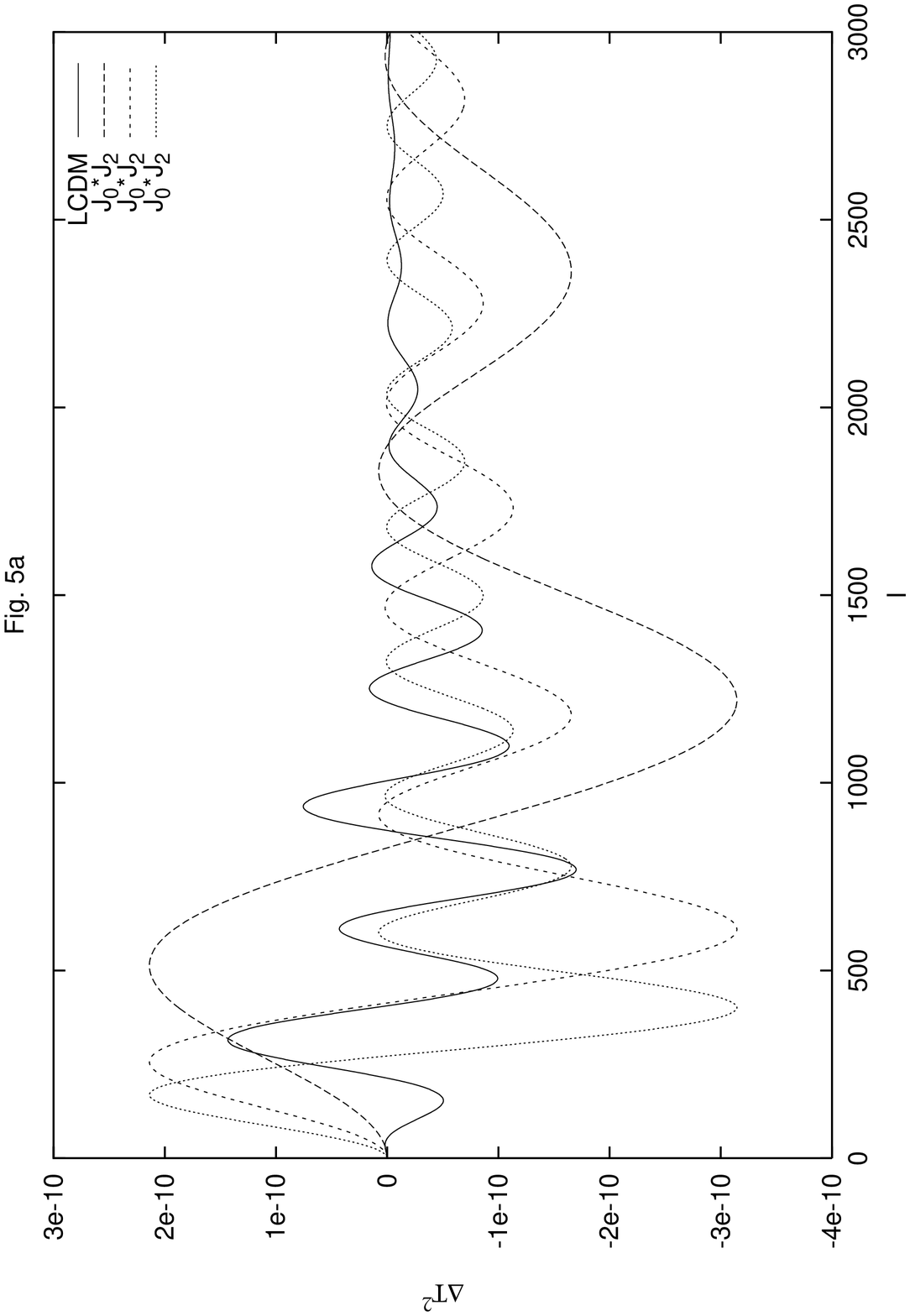} 
\caption{The $T-E$ correlation spectrum $l(l+1)C_{TE,l}/2\pi$
(solid line) for the $\Lambda$CDM
cosmology.   Superposed on the correlation
spectrum are the (arbitrarily scaled)
$k$-space filter functions $J_2(kR)J_0(kR)$ for 
some choices of $R$'s. 
Hardly can any choice of $R$ yield $k$-space 
filter functions that oscillate in phase with the correlation
spectrum.}
\end{figure}

\clearpage
\begin{figure}
\plotone{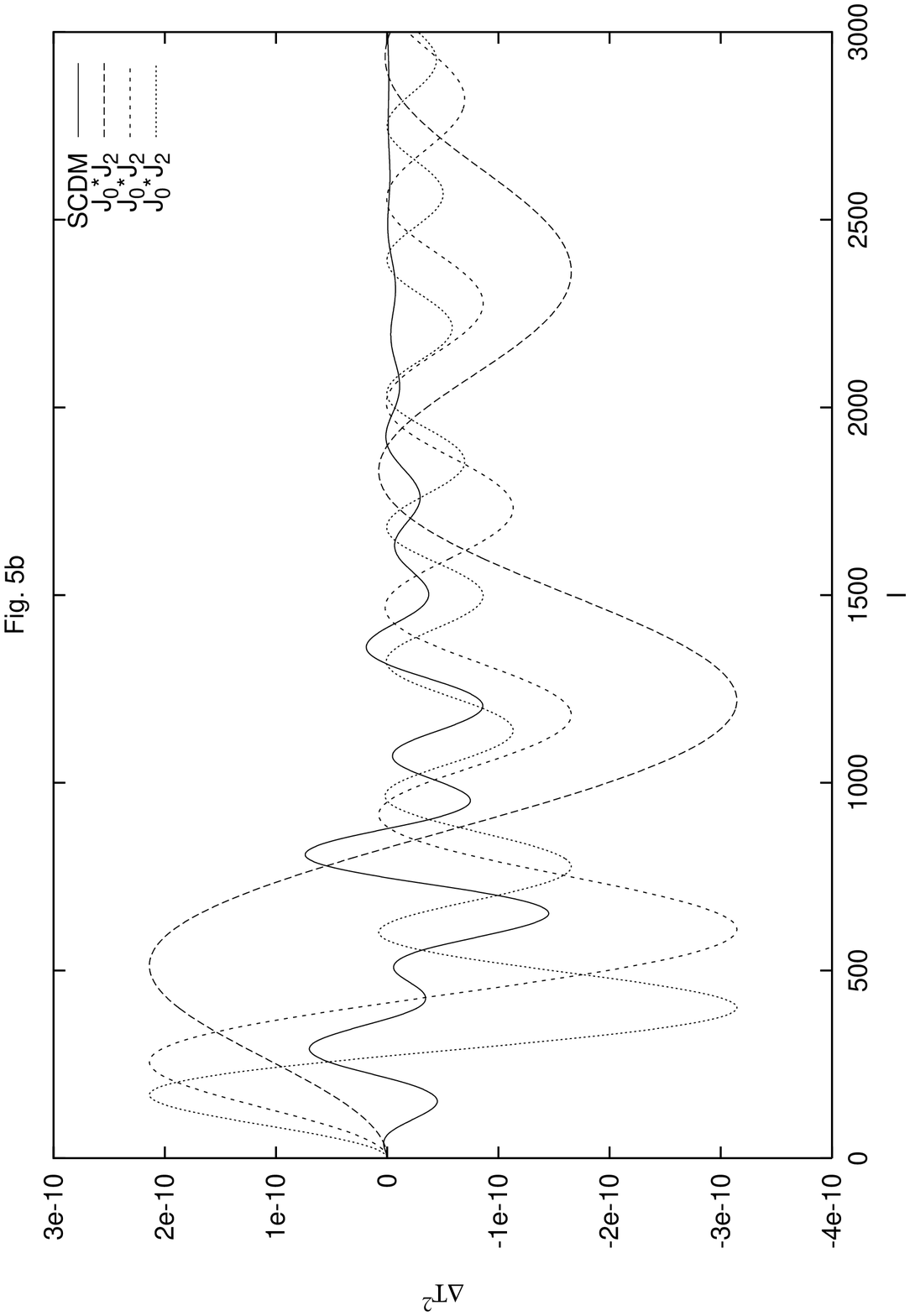} 
\caption{Same as Fig.(5a), but for the standard CDM cosmology.}
\end{figure}

\clearpage
\begin{figure}
\plotone{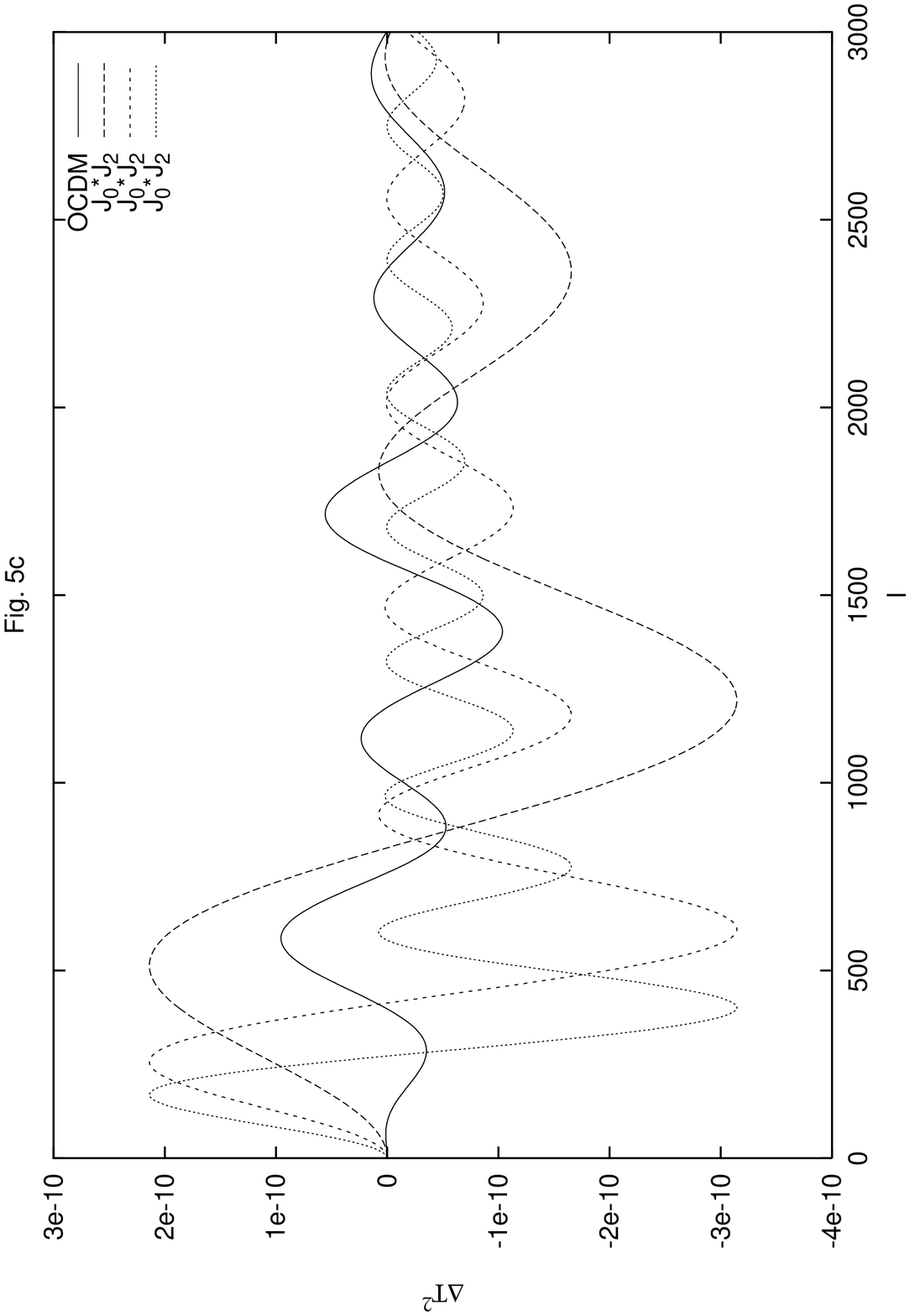} 
\caption{Same as Fig.(5a), but for the open CDM cosmology.}
\end{figure}

\clearpage
\begin{figure}
\plotone{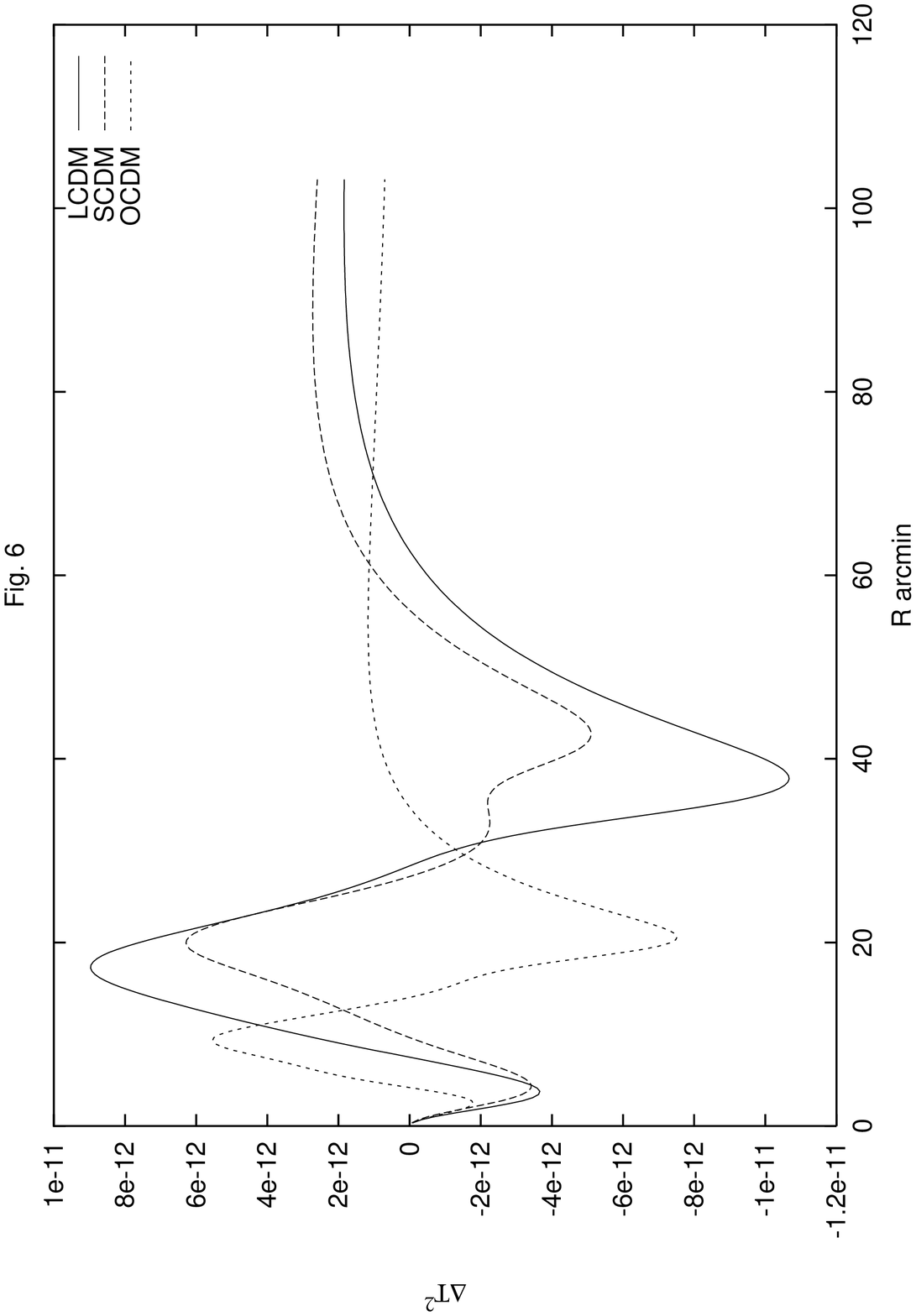} 
\caption{The filtered $T-E$ correlation $\bar C_{TE}$ as functions
of $R$ for the three cosmologies.   The units are the same
as Fig.(2).   The signal strengths are seen to be substantially
lower than the peaks in Fig.(5).}
\end{figure}

\clearpage
\begin{figure}
\plotone{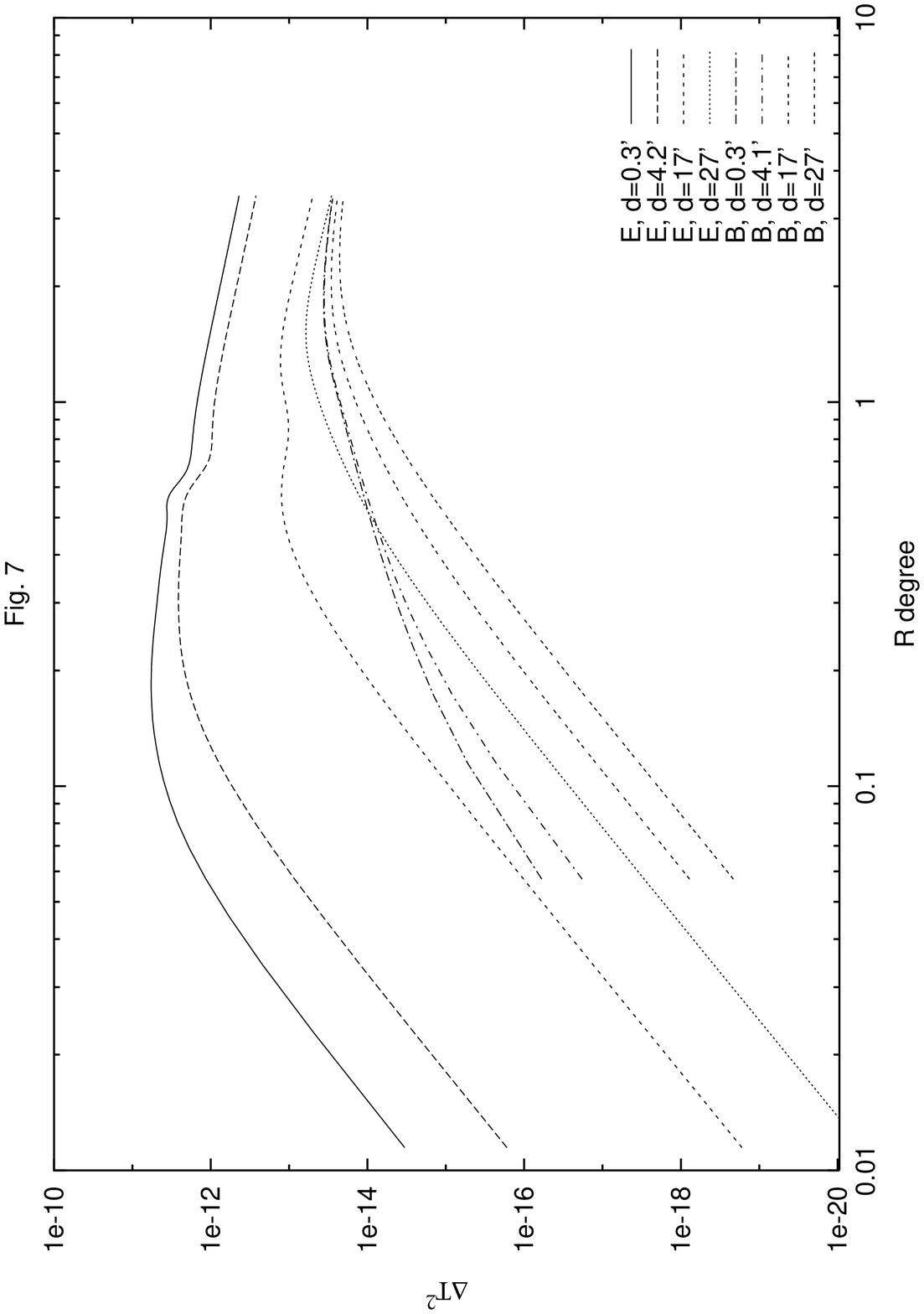} 
\caption{The finite-beam effects on the detected $E$ and $B$-mode surface
brightness as functions of the radius of sky annulus $R$ and of various
half-beam widths $d$ for the fiducial $\Lambda$CDM cosmology.
The reduction of surface brightness is due to the Gaussian
cutoff of finite beam in the window function as indicated in Eq.(22).}
\end{figure}

\clearpage
\begin{figure}
\plotone{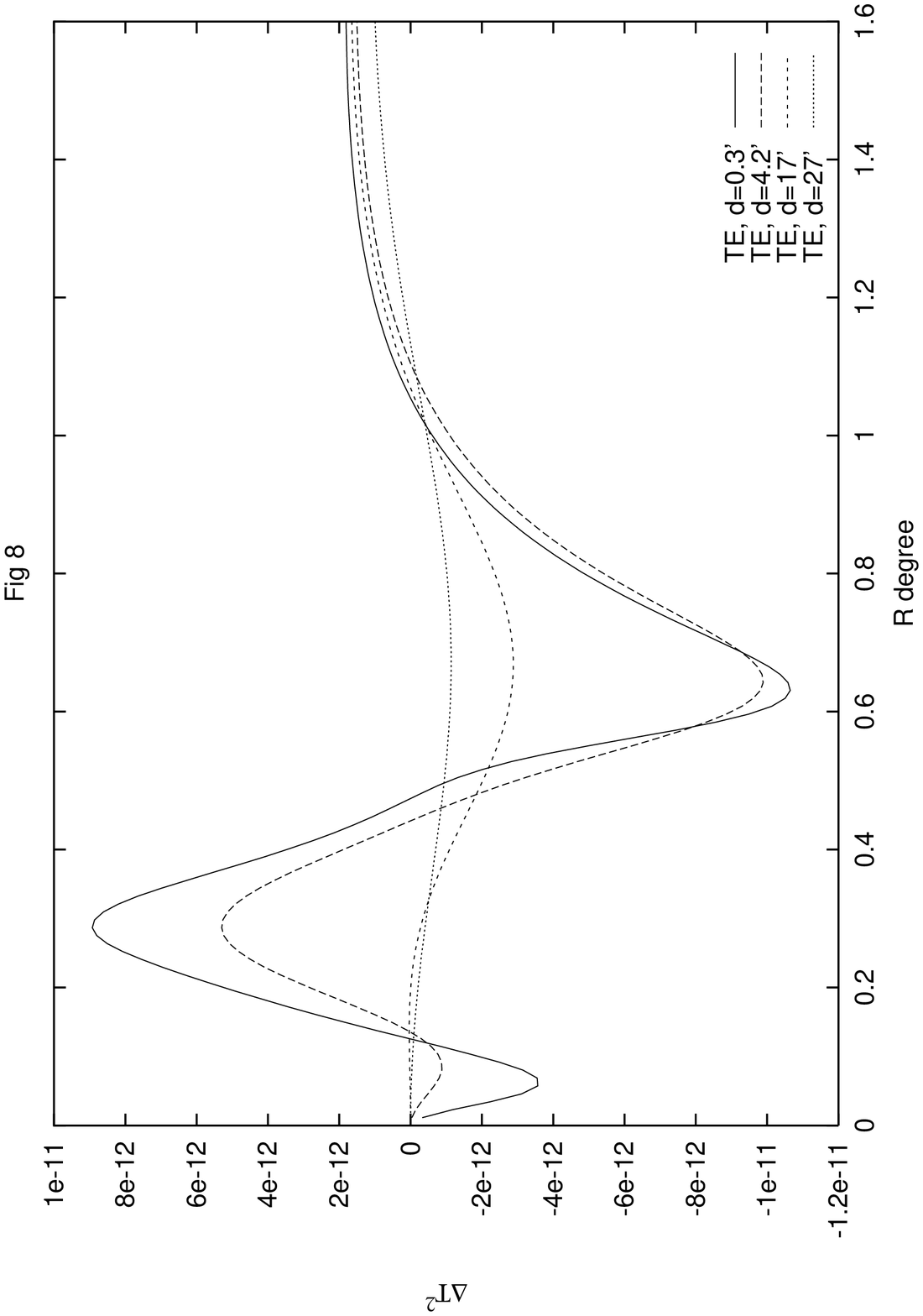} 
\caption{ The finite-beam effects on the detected $T-E$ correlation as functions
of $R$ and of $d$ for the fiducial $\Lambda$CDM cosmology.}
\end{figure}

\end{document}